\documentclass[twocolumn,prb,10pt,aps,floatfix]{revtex4-1}
\usepackage[english]{babel}
\usepackage{graphicx}
\usepackage[pdftex]{epsfig}
\usepackage{ragged2e}
\usepackage{amsmath,amssymb}
\usepackage{times}
\usepackage[colorlinks,linkcolor=blue,citecolor=blue,urlcolor=blue]{hyperref}
\usepackage{color}
\usepackage[table,xcdraw]{xcolor}
\usepackage{epstopdf}
\usepackage{physics}
\usepackage[export]{adjustbox}
\usepackage{dsfont}

\newcommand{\be}{\begin{equation}}
\newcommand{\ee}{\end{equation}}

\newcommand{\bee}{\begin{equation*}}
\newcommand{\eee}{\end{equation*}}

\newcommand{\1}{\hspace*{-1pt}}
\newcommand{\2}{\hspace*{-2pt}}

\begin{document}
\title{Effects of two-loop contributions in the pseudofermion functional renormalization group method for quantum spin systems}
\author{Marlon R\"uck$^{1}$}
\author{Johannes Reuther$^{1, 2}$}
\affiliation{$^1$Dahlem Center for Complex Quantum Systems and Institut f\"ur Theoretische Physik, Freie Universit\"{a}t Berlin, Arnimallee 14, 14195 Berlin, Germany}
\affiliation{$^2$Helmholtz-Zentrum f\"{u}r Materialien und Energie, Hahn-Meitner-Platz 1, 14019 Berlin, Germany} 

\date{\today}

\begin{abstract}
We implement an extension of the pseudofermion functional renormalization group (PFFRG) method for quantum spin systems that takes into account two-loop diagrammatic contributions. An efficient numerical treatment of the additional terms is achieved within a nested graph construction which recombines different one-loop interaction channels. In order to be fully self consistent with respect to self-energy corrections we also include certain three-loop terms of Katanin type. We first apply this formalism to the antiferromagnetic $J_1$-$J_2$ Heisenberg model on the square lattice and benchmark our results against the previous one-loop plus Katanin approach. Even though the RG equations undergo significant modifications when including the two-loop terms, the magnetic phase diagram -- comprising N\'eel ordered and collinear ordered phases separated by a magnetically disordered regime -- remains remarkably unchanged. Only the boundary position between the disordered and the collinear phases is found to be moderately affected by two-loop terms. On the other hand, critical RG scales, which we associate with critical temperatures $T_\text{c}$, are reduced by a factor of $\sim2$ indicating that the two-loop diagrams play a significant role in enforcing the Mermin-Wagner theorem. Improved estimates for critical temperatures are also obtained for the Heisenberg ferromagnet on the 3D simple cubic lattice where errors in $T_\text{c}$ are reduced by $\sim34\%$.
\end{abstract}
\maketitle

\section{Introduction}\label{introduction}
One of the most fascinating situations in quantum magnetism arises when the effects of small spin magnitudes, low dimensional lattices and frustrating interactions cooperate, to melt conventional magnetic long-range order in the ground state. This may result in a variety of different spin states, ranging from valence-bond crystals\cite{frustration_book} that still exhibit a ''hidden`` type of spontaneous symmetry breaking to quantum spin liquids\cite{anderson73,balents10,savary16,lee08} which are characterized by topological order\cite{wen91,wen02} and fractional quasiparticle excitations\cite{read91,wen02}. Spin liquids can again appear in many different flavors as they may have chiral\cite{wen89,kalmeyer87,schroeter07,yao07} or nematic\cite{andreev84,gorkov90,podolsky05,iqbal16} properties and may be described by various different types of effective lattice-gauge theories\cite{wegner71,senthil00,rantner02,hermele04}.

Even though such scenarios were considered very exotic in the times of their first proposal by Anderson in 1973\cite{anderson73}, the rise of powerful numerical approaches in the past decades has given convincing evidence that the general phenomenon of frustration-induced destruction of ground-state magnetic order is actually not rare in dimensions $D\geq 2$. Indeed, for antiferromagnetic spin-1/2 Heisenberg models, non-magnetic phases have been proposed on all standard 2D lattices, i.e., square\cite{jiang12,schulz96,isaev09,darradi08,gong14,richter10,poilblanc17,wang17,wang16,hu13,singh99,kotov99,reuther10}, triangular\cite{zhu15,hu15,iqbal16_2}, honeycomb\cite{mulder10,albuquerque11,bishop12,gong13}, and kagome\cite{jiang08,yan11,iqbal13,he17} lattices -- at least when frustrating first {\it and} second neighbor interactions are considered. There is also growing numerical evidence that the effects of frustration may even stabilize magnetically disordered states in 3D spin systems, such as Heisenberg models on simple cubic\cite{iqbal16_3,farnell16}, pyrochlore\cite{berg03,tsunetsugu01} and hyperkagome\cite{lawler08,buessen16} lattices.

Despite the recent progress in the development of new numerical approaches, the unambiguous identification of ground state properties of generic quantum spin models still represents a serious challenge and often requires severe approximations. This is because any numerical ground-state solver needs to correctly capture the non-trivial interplay between magnetic long-range order and quantum fluctuations, both of which are complicated many-body effects. Approaches such as exact diagonalization, DMRG\cite{white92,schollwock05}, iPEPS\cite{jordan08}, coupled cluster method\cite{zeng98}, and quantum Monte Carlo\cite{ceperley77,sandvik91} have been successfully applied to quantum spin models, however, they are all limited in certain respects. For example, DMRG has become very powerful even in 2D, but its application in 3D seems to be out of reach. On the other hand, exact diagonalization is independent of the lattice dimension but restricted to very small system sizes. Quantum Monte Carlo stands out in the sense that in non-frustrated (i.e., sign-problem free) cases numerical errors are only of statistical type, however, the limitation to non-frustrated systems excludes many interesting models.

The PFFRG method\cite{reuther10,reuther11,iqbal15,iqbal16_3,buessen16,reuther11_2,iqbal16,hering17,baez17} is another numerical approach which has recently proven to be applicable to the ground state properties of quantum spin systems. Following a fermionic reformulation of the spin degrees of freedom, the system is treated within the established functional renormalization group (FRG) technique\cite{wetterich93,metzner12,platt13}, which sums up diagrammatic vertex contributions in different one-loop interaction channels. Apart from so-called Katanin terms\cite{katanin04} which effectively act as fermionic self-energy corrections, two-loop contributions have been neglected so far. Already at this level of approximation, the PFFRG turns out to be surprisingly powerful and flexible, as it allows to treat arbitrary lattices in 2D and 3D\cite{reuther10,reuther11,iqbal15,iqbal16_3,buessen16}, isotropic and anisotropic\cite{reuther11_2,iqbal16,hering17} two-body interactions, unrestricted spin magnitudes\cite{baez17} $S$ as well as arbitrary frustrated interactions. However, since the PFFRG performs diagrammatic summations in a situation where a small parameter is typically absent, the errors associated with the neglected contributions are {\it a priori} very hard to estimate. Important insight in this context is gained by noting that to leading order, the one-loop PFFRG is separately exact in the large $S$ limit\cite{baez17} and in the large $N$ limit\cite{buessen17,roscher17} -- at least until the point where an instability occurs during the RG flow [here $N$ refers to a generalization of the spins' symmetry group to SU(N)]. Consequently, PFFRG can be expected to correctly capture the ground state properties of spin systems deep in magnetically ordered phases (where a large $S$ description applies) and deep in magnetically disordered phases such as spin liquids or valence-bond crystals (where a large $N$ description applies). However, close to quantum critical points, subleading two-loop contributions may become important such that the exact positions of phase boundaries might still be subject to errors in the PFFRG.

In this article, we study a two-loop PFFRG approach to investigate the effects of diagrammatic contributions beyond one-loop, and to find out to what extend the results of previous PFFRG schemes are already converged. To this end, we implement a formalism that is closely related to the one proposed by A. Eberlein\cite{eberlein14}, which recasts two-loop contributions into an effective one-loop form and which is exact up to the third order in the effective interaction. As detailed below, in order to guarantee the self-consistent treatment of self-energy renormalization effects, our scheme even involves certain three-loop Katanin-type contributions that have not been considered in Ref.~\onlinecite{eberlein14}.

As a prototypical frustrated spin system to test our approach, we first consider the Heisenberg model on the square lattice with antiferromagnetic first and second neighbor interactions $J_1$ and $J_2$, respectively. The overall sequence of ground-state phases of this model is well known: As a function of the parameter $g=J_2/J_1$ the system first shows antiferromagnetic N\'eel order for $0\leq g \leq g_{\text{c}1}$, where most numerical methods find $g_{\text{c}1}$ to be in the range $0.4<g_{\text{c}1}<0.5$\cite{jiang12,isaev09,darradi08,gong14,wang17,hu13,singh99}. For comparison, a previous one-loop (plus Katanin) PFFRG study\cite{reuther10} found $g_{\text{c}1}\approx 0.4...0.45$. Increasing $g$ beyond $g_{\text{c}1}$, magnetic long-range order is destabilized due to the frustration effect and the system resides in a magnetically disordered phase. Despite intense numerical research for more than two decades, the exact nature of this intermediate phase is still under debate with suggestions ranging from different types of valence-bond solids\cite{isaev09,darradi08,singh99,kotov99} to quantum spin liquids\cite{jiang12,gong14,wang17,wang16,hu13}. For larger $g\geq g_{\text{c}2}$ the system again shows magnetic long-range order of so-called collinear type where the spins align antiferromagnetically in one lattice direction and ferromagnetically in the other. Most numerical methods find $0.6<g_{\text{c}2}<0.66$\cite{jiang12,isaev09,gong14,richter10,singh99,kotov99} while the PFFRG study in Ref.~\onlinecite{reuther10} obtained $g_{\text{c}2}\approx 0.66...0.68$.

A central result of our study is that the above-mentioned sequence of quantum phases remains unchanged when adding two-loop contributions, with only small shifts of the phase boundary $g_{\text{c}2}$. This shift reduces the extend of the intermediate non-magnetic phase, to better agree with other numerical methods. In total, this finding indicates that already on the one-loop level, PFFRG phase diagrams can be expected to be mostly converged and to give good estimates of phase boundaries. On the other hand, the added two-loop terms are found to have a large effect on critical temperatures $T_\text{c}$ and the fulfillment of the Mermin-Wagner theorem. An analysis of critical temperatures is performed as in Ref.~\onlinecite{iqbal16_3}, where the RG scale $\Lambda_\text{c}$ at which the flow runs into a magnetic instability is proportional to $T_\text{c}$. In magnetically ordered 2D Heisenberg systems, PFFRG typically predicts a finite $T_c$ on the order of the exchange couplings, in strong violation to the Mermin-Wagner theorem (according to which critical temperatures should be suppressed to zero in spin-isotropic 2D systems due to strong infrared thermal fluctuations\cite{mermin66,hohenberg67}). We find that for the $J_1$-$J_2$ square lattice Heisenberg model the added two-loop contributions reduce $T_\text{c}$ by a factor of two and, therefore, lead to a significantly better fulfillment of the Mermin-Wagner theorem. We also discuss the precise form of the diagrammatic contributions which are responsible for this improvement.

To complete the analysis of critical temperatures, we additionally consider the 3D Heisenberg case where $T_\text{c}$ is typically finite. As an example, we discuss one-loop and two-loop PFFRG results for the 3D nearest neighbor ferromagnetic Heisenberg model on the simple cubic lattice. For the established one-loop PFFRG scheme, we find that $T_c$ is overestimated by $\sim29\%$ as compared to (quasi-) exact quantum Monte Carlo results. As an effect of two-loop contributions, this error is reduced to an overestimation of $\sim19\%$, which is a further indication that critical temperatures come out substantially improved due to the additional terms.

The paper is structured as follows: The method section~\ref{sec:method} first reviews the general PFFRG formalism and explains the Katanin truncation scheme (Sec.~\ref{sec:oneloop}). The two-loop extension (including the Katanin-corrected terms) and its diagrammatic implementation is discussed in the following Section~\ref{sec:twoloop}. After some remarks about the numerical evaluation of the RG equations (Sec.~\ref{sec:numerics}) we present the results of our study in Sec.~\ref{sec:results}. We first investigate the $J_1$-$J_2$ square lattice Heisenberg model (Sec.~\ref{sec:j1j2}) followed by a brief discussion of the 3D simple cubic Heisenberg model (Sec.~\ref{sec:sc_fm}). The paper ends with a conclusion in Section~\ref{sec:conclusion}. Three appendices contain further details about the two-loop scheme such as the derivation of the RG equations (Appendix~\ref{app1}), the diagrammatic discussion of the Mermin-Wagner theorem (Appendix~\ref{app2}) and the numerical implementation of the $\Lambda$ integration (Appendix~\ref{app3}).

\section{Method}\label{sec:method}
\subsection{General PFFRG scheme and Katanin truncation}
\label{sec:oneloop}
Before we discuss the implementation of two-loop terms, we first briefly review the general PFFRG scheme and the Katanin truncation as it has been applied previously\cite{reuther10}. We start with a generic Heisenberg model of the form
\begin{equation}
H=\sum_{(ij)}J_{ij} \mathbf{S}_i \mathbf{S}_j\;,\label{ham_generic}
\end{equation}
where $i$, $j$ are the sites of an arbitrary lattice and $J_{ij}$ can be any set of exchange couplings between sites $i$ and $j$. The sum runs over pairs of sites $(ij)$. Within all PFFRG approaches, the spin operators are first recast into a fermionic form, using
\begin{equation}
S^{\mu}_i=\frac{1}{2}\sum\limits_{\alpha,\beta} f^{\dagger}_{i\alpha}\sigma^{\mu}_{\alpha\beta}f_{i\beta}\;.\label{pseudofermions}
\end{equation}
Here, $f_{i\alpha}$ ($f^{\dagger}_{i\alpha}$) are spinful fermionic annihilation (creation) operators with $\alpha=\uparrow,\downarrow$ acting on site $i$. Furthermore, $\sigma^{\mu}_{\alpha\beta}$ ($\mu \in \{x,y,z\}$) denotes the Pauli matrices. The representation in Eq.~(\ref{pseudofermions}) needs to be treated with some caution as it introduces unphysical spin-zero states (with local occupations $Q_i\equiv f^{\dagger}_{i\uparrow}f_{i\uparrow}+f^{\dagger}_{i\downarrow}f_{i\downarrow}=0$ or $Q_i=2$) in addition to the physical spin-1/2 states (with local occupation $Q_i=1$). A convenient method to eliminate possible unwanted contributions of the $S=0$ states in the PFFRG results, is to add a local level repulsion term $-A \sum_i\mathbf{S}_i^2$ to the Hamiltonian\cite{baez17}. If $A$ is positive, the energy levels in the physical spin-1/2 subspace are shifted down compared to the unphysical states which guarantees that for $A$ sufficiently large, unphysical states do not contribute to the ground-state properties. We note, however, that for generic Heisenberg models -- including the systems studied here -- there is no qualitative change in the results when increasing $A$ from zero, which indicates that already for $A=0$, unphysical states do not contribute (for a detailed discussion of this important point, see Ref.~\onlinecite{baez17}). This property can be understood by noting that an unphysically occupied $S=0$ site acts like a magnetic vacancy in the spin lattice, which costs an excitation energy on the order of the exchange couplings. The ground state without level-repulsion terms should therefore not be poisoned with contributions from unphysical fermionic occupations. All results presented in the following are calculated for $A=0$.

Next, the fermionic model obtained when inserting Eq.~(\ref{pseudofermions}) into Eq.~(\ref{ham_generic}) is treated within the standard functional renormalization group (FRG) framework\cite{wetterich93,metzner12,platt13}. A somewhat unusual situation occurs because the fermionic system is purely quartic in the fields, without any kinetic hopping terms. As a consequence, the free fermionic propagator $G_0$ on the imaginary Matsubara axis has the simple form
\begin{equation}
G_0(\omega)=\frac{1}{i\omega}\;,\label{bare_g}
\end{equation}
and is local in real space to all orders of diagrammatic expansions.

The first important step in all FRG schemes is to regularize the free propagator. Within PFFRG this amounts to introducing an artificial Heaviside-step function in $G_0$ which suppresses the fermionic propagation in the infrared limit, i.e., we replace
\begin{equation}
G_0(\omega)\rightarrow G_0^{\Lambda}(\omega)=\theta\left ( \left | \omega  \right | - \Lambda \right) G_0(\omega)\;, \label{cutoff}
\end{equation}
where $\Lambda$ is the so-called RG scale. Formally, this regularization connects the trivial limit $\Lambda\rightarrow\infty$, where the propagator vanishes identically and only bare interactions $J_{ij}$ remain, with the fully renormalized and physically relevant limit $\Lambda=0$. The FRG describes the system's evolution between both limits in terms of flow equations for the one-particle irreducible $m$-particle vertex functions. These equations are formally exact and can be derived from the scale derivative of the effective action. The first two equations for the self energy $\Sigma^{\Lambda}$ and the two-particle vertex $\Gamma^{\Lambda}$ read
\begin{align}
 \frac{d}{d\Lambda}\Sigma^{\Lambda}\left(1\right)&=-\frac{1}{2\pi}\sum\limits_{2}\Gamma^{\Lambda}\left(1,2;1,2\right)S^{\Lambda}\left(2\right)\;,\label{frg_eq_first}\\ 
 \frac{d}{d\Lambda}\Gamma^{\Lambda}\left(1',2';1,2\right)&=\frac{1}{2\pi}\sum\limits_{3,4}\Big[ \Gamma^{\Lambda}\left( 1',2';3,4\right) \Gamma^{\Lambda}\left( 3,4;1,2\right) \nonumber \\
&-\Gamma^{\Lambda}( 1',4;1,3) \Gamma^{\Lambda}( 3,2';4,2) \1 - \1 (3 \leftrightarrow 4 ) \nonumber \\
&+\Gamma^{\Lambda}( 2',4;1,3) \Gamma^{\Lambda}( 3,1';4,2) \1 + \1 (3 \leftrightarrow 4 ) \1\Big ] \nonumber \\ &\times G^{\Lambda}(3)S^{\Lambda}(4)\notag\\
&+\frac{1}{2\pi}  \sum\limits_{3} \Gamma_3^\Lambda \left( 1',2',3;1,2,3\right) S^{\Lambda}\left(3\right)\,.\label{frg_eq_second} 
\end{align}
Here, arguments ``1'' denote multi-indices comprising the Matsubara frequency, lattice site, and spin index, i.e., $1=\{\omega_1,i_1,\alpha_1\}$. Furthermore, $\Gamma^\Lambda_3$ stands for the three-particle vertex. The FRG equations contain the fully dressed propagator
\begin{equation}
G^{\Lambda}=\left[ \left ( G^{\Lambda}_0 \right )^{-1} -\Sigma^{\Lambda}\right ]^{-1}\;,\label{dressed_g}
\end{equation}
and the so-called single scale propagator
\begin{equation}
 S^{\Lambda}=G^{\Lambda} \frac{d}{d\Lambda}\left[ G^{\Lambda}_0 \right ]^{-1} G^{\Lambda}\;,\label{singlescale}
\end{equation}
where the latter follows from a $\Lambda$-derivative of $G^{\Lambda}$, acting only on the $\Lambda$-dependence contained in $G^{\Lambda}_0$ but not on $\Sigma^{\Lambda}$. Note that for Heisenberg systems on lattices with equivalent sites (particularly, for lattices with a mono-atomic unit cell) the propagators are independent of spin and site indices, i.e., $G^\Lambda(1)\equiv G^\Lambda(\omega_1)$ and $S^\Lambda(1)\equiv S^\Lambda(\omega_1)$. Similar flow equations can also be formulated for higher vertices where the $\Lambda$-derivative of each $m$-particle vertex is determined by all $m'$-particle vertices with $m'\leq m+1$. In total, this result in an infinite but exact hierarchy of coupled FRG differential equations.
 \begin{figure*}[t]
\includegraphics[width=0.99\linewidth]{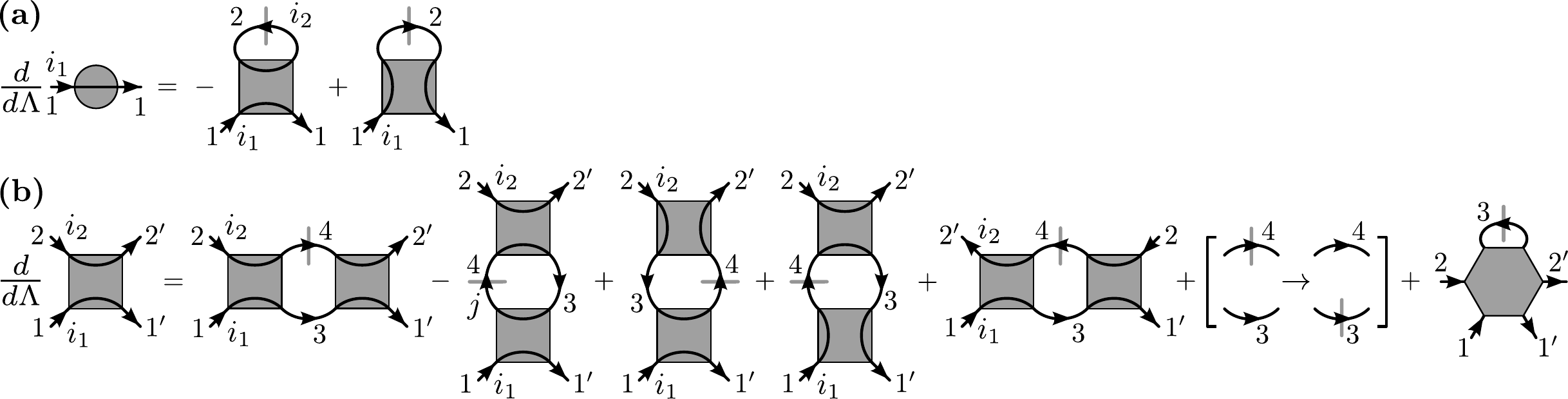}
\caption{Diagrammatic illustration of the PFFRG equations for (a) the self energy $\Sigma^\Lambda$ and (b) the two-particle vertex $\tilde{\Gamma}^\Lambda$, where the terms appear in the same order as in Eqs.~(\ref{FRG_sigma}) and (\ref{FRG_gamma}). The vertex functions $\Sigma^\Lambda$, $\tilde{\Gamma}^\Lambda$, and $\Gamma_3^\Lambda$ are represented by gray shaded disks, squares, and hexagons, respectively. The single-scale propagator $S^\Lambda$ (fully dressed propagator $G^{\Lambda}$) is drawn as an arrow with (without) a slash. Numbers $1$, $1',\ldots$ stand for frequency and spin variables while site indices are specified by $i_1$, $i_2$, $j$ (note the site variables do not change along fermion lines). The square bracket in (b) indicates that the previous five terms additionally appear with the single-scale propagator occurring on the fermion line 3 [which corresponds to the term $\sim G^\Lambda(\omega_4)S^\Lambda(\omega_3)$ in Eq.~(\ref{FRG_gamma})].\label{fig:diag_frg_eq}}
\end{figure*}

The above FRG equations can be written in a more convenient form which better highlights the spatial site-index structure of the different terms. The locality of the propagator $G_0^\Lambda$ implies that a two-particle vertex $\Gamma^\Lambda(1',2';1,2)$ cannot change its site index along fermion lines and, hence, $\Gamma^\Lambda(1',2';1,2)$ can only depend on two sites, with either $i_{1'}=i_1$, $i_{2'}=i_2$ or $i_{1'}=i_2$, $i_{2'}=i_1$. Taking into account the antisymmetry of fermionic vertices under the exchange of two external variables, one may therefore parametrize the site dependence of $\Gamma^\Lambda(1',2';1,2)$ by
\begin{align}
\Gamma^\Lambda(1',2';1,2)&=\tilde{\Gamma}^\Lambda_{i_1 i_2}(1',2';1,2)\delta_{i_{1'}i_1}\delta_{i_{2'}i_2}\notag\\
&-\tilde{\Gamma}^\Lambda_{i_2 i_1}(1',2';2,1)\delta_{i_{1'}i_2}\delta_{i_{2'}i_1}\;.\label{parametrize_site}
\end{align}
Note that the new vertex $\tilde\Gamma^\Lambda$ obeys $\tilde\Gamma_{i_1i_2}^\Lambda(1',2';1,2)=\tilde\Gamma_{i_2i_1}^\Lambda(2',1';2,1)$ and that multi-indices ``1'' in the arguments of $\tilde\Gamma^\Lambda$ only contain the frequency $\omega_1$ and the spin $\alpha_1$ while the site dependencies are written as a subscript index. Inserting the parametrization of Eq.~(\ref{parametrize_site}) into Eqs.~(\ref{frg_eq_first}) and (\ref{frg_eq_second}) one obtains
\begin{align}
\frac{d}{d\Lambda}\Sigma^{\Lambda}\left(\omega_1\right)&=\frac{1}{2\pi}\sum\limits_{2}\Big[-\sum_j\tilde\Gamma_{i_1j}^{\Lambda}(1,2;1,2)\notag\\
&+\tilde\Gamma_{i_1i_1}^{\Lambda}(1,2;2,1)\Big] S^{\Lambda}(\omega_2)\;,\label{FRG_sigma}
\end{align}
\begin{align}
\frac{d}{d\Lambda}\tilde\Gamma_{i_1i_2}^{\Lambda}(1',2';1,2)&=\frac{1}{2\pi}\2\sum_{3,4}\2\Big[\tilde\Gamma_{i_1i_2}^{\Lambda}\1(\1 1'\1 ,\1 2';\1 3,\1 4)\tilde\Gamma_{i_1i_2}^{\Lambda}\1(3,\1 4;\1 1,\1 2)\notag\\
&-\sum_j\tilde\Gamma_{i_1j}^{\Lambda}(1',4;1,3)\tilde\Gamma_{ji_2}^{\Lambda}(3,2';4,2)\notag\\
&+\tilde\Gamma_{i_1i_2}^{\Lambda}(1',4;1,3)\tilde\Gamma_{i_2i_2}^{\Lambda}(3,2';2,4)\notag\\
&+\tilde\Gamma_{i_1i_1}^{\Lambda}(1',4;3,1)\tilde\Gamma_{i_1i_2}^{\Lambda}(3,2';4,2)\notag\\
&+\tilde\Gamma_{i_1i_2}^{\Lambda}(3,2';1,4)\tilde\Gamma_{i_1i_2}^{\Lambda}(1',4;3,2)\Big]\notag\\
&\times \Big[G^\Lambda(\omega_3)S^\Lambda(\omega_4)+G^\Lambda(\omega_4)S^\Lambda(\omega_3)\Big]\notag\\
&+\frac{1}{2\pi}  \sum\limits_{3} \Gamma_3^\Lambda \left( 1',2',3;1,2,3\right) S^{\Lambda}\left(\omega_3\right)\,.\label{FRG_gamma}
\end{align}
A diagrammatic representation of these equations is shown in Fig.~\ref{fig:diag_frg_eq}. Lines one to five of Eq.~(\ref{FRG_gamma}) contain contributions from different interaction channels, which can be distinguished by their real-space index structure. This channel decomposition will turn out to be useful for the construction of two-loop terms in the next section. Also, the five terms are associated with different physical properties of spin phases. The first line in Eq.~(\ref{FRG_gamma}) is the particle-particle ladder which describes fermionic pairing effects and which is essential for the description of $\mathds{Z}_2$ spin liquids\cite{wen91}. The second line is the RPA channel which is responsible for the formation of magnetic long-range order. The RPA terms also guarantee that the PFFRG is exact in the large $S$ limit\cite{baez17}. While the vertex correction terms in lines three and four cannot be attributed to a particular spin limit, the fifth line contains the particle-hole ladder which ensures the exactness in the large $N$ limit\cite{affleck88,buessen17,roscher17}. This term describes fluctuations in the fermionic hopping channel which together with the pairing channel is important for the formation of non-magnetic states\cite{wen02}.

For a numerical treatment of the PFFRG equations, the two-particle vertex needs to be further parametrized in its spin and frequency arguments. Particularly, $\tilde{\Gamma}$ can be written as a sum of a spin-spin interaction vertex $\tilde\Gamma_\text{s}^\Lambda$ and a density-density interaction vertex $\tilde\Gamma_\text{d}^\Lambda$,
\begin{align}
\tilde\Gamma^{\Lambda}_{i_1 i_2}(1',2';1,2) & = \Big [ \tilde\Gamma^{\Lambda}_{\text{s} \, i_1 i_2}\left( \omega_1',\omega_2';\omega_1,\omega_2\right)\sum_\mu\sigma^{\mu}_{\alpha_{1'} \alpha_{1}}\sigma^{\mu}_{\alpha_{2'} \alpha_{2}} \nonumber \\ &+\tilde\Gamma^{\Lambda}_{\text{d} \, i_1 i_2}\left( \omega_1',\omega_2';\omega_1,\omega_2\right)\delta_{\alpha_{1'} \alpha_{1}}\delta_{\alpha_{2'} \alpha_{2}} \Big ] \nonumber \\ &\times \delta(\omega_1+\omega_2-\omega_{1'}-\omega_{2'})\;.\label{parametrize}
\end{align}
In this formulation, the initial conditions of the RG flow defined at $\Lambda\rightarrow\infty$ are given by $\tilde{\Gamma}_{\text{s} \, i_1 i_2}^\infty=J_{i_1 i_2}/4$, $\tilde{\Gamma}_{\text{d} \, i_1 i_2}^\infty=0$, and $\Sigma^\infty=0$. As it only complicates the equations, we will not make explicit use of this parametrization in the following but discuss the flow equations on the basis of Eqs.~(\ref{FRG_sigma}) and (\ref{FRG_gamma}).
 \begin{figure*}[t]
\includegraphics[width=0.99\linewidth]{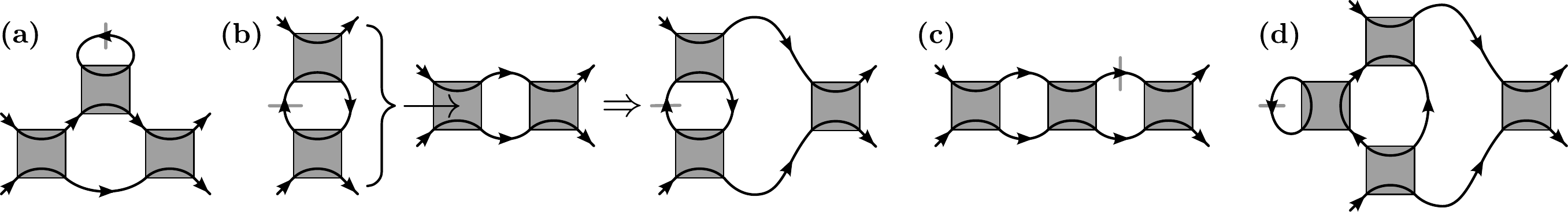}
\caption{(a) Example for an additional three-particle contribution that is generated within the Katanin truncation [Eq.~(\ref{katanin})]. The depicted graph is obtained by inserting a self energy correction into the particle-particle interaction channel which results in a three-particle term of the form $\sim\sum_3 \Gamma_3^\Lambda (1',2',3;1,2,3) S^{\Lambda}(\omega_3)$ [last term in Eq.~(\ref{FRG_gamma}) and in Fig.~\ref{fig:diag_frg_eq}(b)]. (b) Example for the construction of a nested graph in the two-loop extension beyond the Katanin truncation. Two interaction channels are inserted into each other where in one of them (the ``outer'' graph) the single-scale propagator $S^\Lambda$ is replaced by the propagator $G^\Lambda$. (c) The recombination of graphs (here, two particle-particle diagrams) can also lead to terms which are not of three-particle type, see text for details. (d) An improved level of approximation is obtained when equipping the nested diagrams of the two-loop extension with a Katanin correction. The illustrated graph follows from (b) after a self-energy insertion in the single-scale propagator. \label{fig:nested_diagrams}}
\end{figure*}

On a pure one-loop level, the three-particle term $\sim\Gamma^\Lambda_3$ in Eq.~(\ref{FRG_gamma}) is neglected completely which immediately leads to a closed set of flow equations. This approximation, however, turns out to be insufficient to correctly describe the ground-state phases of quantum spin models; particularly non-magnetic states cannot be captured. A crucial improvement comes with the so-called Katanin truncation\cite{katanin04} which lies at the heart of the PFFRG approach as it has been applied previously. Within this scheme, the three particle term in Eq.~(\ref{FRG_gamma}) is also ignored but the single-scale propagator $S^\Lambda$ in this equation is replaced by
\begin{equation}
S^{\Lambda}\longrightarrow -\frac{d}{d\Lambda}G^{\Lambda}=S^{\Lambda}-\left(G^{\Lambda}\right)^2\frac{d}{d\Lambda}\Sigma^{\Lambda}\;.\label{katanin}
\end{equation}
The additional term $\sim d\Sigma^\Lambda/d\Lambda$ on the right-hand side of Eq.~(\ref{katanin}) is a self-energy correction which enters the two-particle vertex flow. Since the new single scale propagator is given by the {\it full} $\Lambda$-derivative of $G^\Lambda$, the Katanin scheme guarantees the complete (maximal) feedback of the self energy into the two-particle vertex. The (imaginary) self energy corresponds to a pseudofermion lifetime which describes a reduction of the local magnetic moment due to quantum fluctuations. The full and self-consistent inclusion of self-energy corrections in the flow of the two-particle vertex is, hence, essential for detecting non-magnetic phases in the PFFRG. Without these additional Katanin terms, phase diagrams largely resemble the corresponding classical ones.

Effectively, the new self-energy corrections from the last term on the right hand side of Eq.~(\ref{katanin}) correspond to certain three-particle contributions of the form $\sim\sum_3 \Gamma_3^\Lambda ( 1',2',3;1,2,3) S^{\Lambda}(\omega_3)$ [see last term in Eq.~(\ref{FRG_gamma}) and in Fig.~\ref{fig:diag_frg_eq}(b)]. To see this, one re-expresses the derivative $d\Sigma^\Lambda/d\Lambda$ in Eq.~(\ref{katanin}) by the right side of the flow equation for the self energy [Eq.~(\ref{FRG_sigma})]. As an example of this reformulation, Fig.~\ref{fig:nested_diagrams}(a) shows the additional Katanin diagram that is obtained when the new single-scale propagator is inserted into the particle-particle channel [first term on the right hand side of Eq.~(\ref{FRG_gamma}) and Fig.~\ref{fig:diag_frg_eq}(b)]. This term has exactly the form of the three-particle contribution in the flow equation for $\tilde\Gamma^\Lambda$ and, therefore, effectively acts as a two-loop correction. Diagrams similar to Fig.~\ref{fig:nested_diagrams}(a) are also obtained when inserting the new single-scale propagator into the other interaction channels. While the Katanin terms are indispensable for capturing the correct ground-state physics of quantum spin models, they can also be implemented with relative ease. This is because the two ingredients for the additional diagrams -- the different interaction channels in Eq.~(\ref{FRG_gamma}) and the $\Lambda$-derivative of the self energy in Eq.~(\ref{FRG_sigma}) -- are already included in a pure one-loop scheme such that a Katanin truncation only requires the recombination of known diagrams.

\subsection{Two-loop extension}
\label{sec:twoloop}
We now discuss an extension of the PFFRG approach described above which takes into account additional two-loop terms. This approach closely resembles the one proposed by A. Eberlein which has been applied to the attractive Hubbard model\cite{eberlein14}. As explain below, our formalism even goes beyond A. Eberlein's scheme since it also includes certain three-loop terms. Here we discuss this extension on a diagrammatic level and rather illustratively while Appendix~\ref{app1} contains a more stringent calculation of the flow equations.

The basic idea behind the two-loop extension shares some similarities with the Katanin truncation. It recombines known diagrammatic contributions to obtain new terms of the three-particle type $\sim\sum_3 \Gamma_3^\Lambda ( 1',2',3;1,2,3) S^{\Lambda}(\omega_3)$ [last term in Eq.~(\ref{FRG_gamma}) and in Fig.~\ref{fig:diag_frg_eq}(b)]. The construction of these diagrams [which is schematically illustrated in Fig.~\ref{fig:nested_diagrams}(b)] requires two terms out of the five interaction channels on the right hand side of Eq.~(\ref{FRG_gamma}) or Fig.~\ref{fig:diag_frg_eq}(b). For the example in Fig.~\ref{fig:nested_diagrams}(b), the RPA channel and the particle-particle channel are chosen. In one of these interaction channels [such as the particle-particle channel in Fig.~\ref{fig:nested_diagrams}(b)] the single-scale propagator $S^\Lambda$ is replaced by the propagator $G^\Lambda$ such that the internal fermion lines are given by a product $G^\Lambda G^\Lambda$. In the other interaction channel, the internal fermion lines $~G^\Lambda S^\Lambda$ are kept unchanged. For the construction of a two-loop term, one two-particle vertex in the graph with internal propagators $G^\Lambda G^\Lambda$ is replaced by the graph with internal propagators $G^\Lambda S^\Lambda$. The resulting diagram [right side of Fig.~\ref{fig:nested_diagrams}(b)] has the desired form of the three-particle term in Eq.~(\ref{FRG_gamma}) and Fig.~\ref{fig:diag_frg_eq}(b). More diagrams of this nested form can be constructed by choosing different combinations of interaction channels and taking into account the two possibilities for selecting the two-particle vertex where the insertion can take place.

Some caution is required when inserting interaction channels into each other, as this may also result in a diagram of the form of Fig.~\ref{fig:nested_diagrams}(c). In this specific example, the particle-particle graph has been inserted into itself. The resulting term, however, is no contribution to $\sim\sum_3 \Gamma_3^\Lambda ( 1',2',3;1,2,3) S^{\Lambda}(\omega_3)$. This becomes obvious when cutting the single-scale propagator line in Fig.~\ref{fig:nested_diagrams}(c) which produces a three-particle graph that is not one-particle irreducible. (The one-particle irreducibility means that a diagram cannot be split into two parts when cutting a single fermion line.) In the current one-particle irreducible implementation of the PFFRG such terms must be discarded.

The criterium specifying which channels may be inserted into each other relies on the so-called transfer frequencies $s$, $t$, $u$, which for a vertex $\tilde\Gamma^\Lambda_{i_1 i_2}(1',2';1,2)$ are defined by $s=\omega_1+\omega_2$, $t=\omega_{1'}-\omega_1$, $u=\omega_{1'}-\omega_2$. The interaction channels in lines one to five of Eq.~(\ref{FRG_gamma}) may be grouped according to the transfer frequencies occurring in the internal fermion lines: Exploiting energy conservation in each diagram one finds $s=\omega_3+\omega_4$ ($u=\omega_3-\omega_4$) in the particle-particle (particle-hole) term while for the contributions in lines two, three, and four of Eq.~(\ref{FRG_gamma}) one has $t=\omega_3-\omega_4$. Following this schemes of distinguishing the different interaction channels, we define the quantities
\begin{widetext}
\begin{equation}
X_{s,i_1 i_2}^\Lambda(1',2';1,2)=\frac{1}{2\pi}\sum_{3,4}\tilde\Gamma_{i_1i_2}^{\Lambda}(1',2';3,4)\tilde\Gamma_{i_1i_2}^{\Lambda}(3,4;1,2)
\Big[G^\Lambda(\omega_3)S^\Lambda(\omega_4)+G^\Lambda(\omega_4)S^\Lambda(\omega_3)\Big]\;,\label{xs}
\end{equation} 
\begin{align}
X_{t,i_1 i_2}^\Lambda(1',2';1,2)&=\frac{1}{2\pi}\sum_{3,4}\Big[-\sum_j\tilde\Gamma_{i_1j}^{\Lambda}(1',4;1,3)\tilde\Gamma_{ji_2}^{\Lambda}(3,2';4,2)
+\tilde\Gamma_{i_1i_2}^{\Lambda}(1',4;1,3)\tilde\Gamma_{i_2i_2}^{\Lambda}(3,2';2,4)\notag\\
&+\tilde\Gamma_{i_1i_1}^{\Lambda}(1',4;3,1)\tilde\Gamma_{i_1i_2}^{\Lambda}(3,2';4,2)\Big]\Big[G^\Lambda(\omega_3)S^\Lambda(\omega_4)+G^\Lambda(\omega_4)S^\Lambda(\omega_3)\Big]\;,\label{xt}
\end{align}
\begin{equation}
X_{u,i_1 i_2}^\Lambda(1',2';1,2)=\frac{1}{2\pi}\sum_{3,4}\tilde\Gamma_{i_1i_2}^{\Lambda}(3,2';1,4)\tilde\Gamma_{i_1i_2}^{\Lambda}(1',4;3,2)
\Big[G^\Lambda(\omega_3)S^\Lambda(\omega_4)+G^\Lambda(\omega_4)S^\Lambda(\omega_3)\Big]\;\label{xu}
\end{equation}
such that $d\tilde\Gamma^\Lambda/d\Lambda=X_s^\Lambda+X_t^\Lambda+X_u^\Lambda+1/(2\pi)\sum\limits_{3} \Gamma_3^\Lambda \left( 1',2',3;1,2,3\right) S^{\Lambda}\left(\omega_3\right)$. The full contribution to the two-particle flow in this two-loop extension, containing all allowed diagrammatic recombinations, is then given by
\begin{align}
&\frac{d}{d\Lambda}\tilde\Gamma_{i_1i_2}^{\Lambda}(1',2';1,2)\Big|_\text{tl}=\frac{1}{2\pi}\sum_{3,4}\Big\{\Big[X_{t,i_1i_2}^{\Lambda}(1',2';3,4)+X_{u,i_1i_2}^{\Lambda}(1',2';3,4)\Big]\tilde\Gamma_{i_1i_2}^{\Lambda}(3,4;1,2)\notag\\
&+\tilde\Gamma_{i_1i_2}^{\Lambda}(1',2';3,4)\Big[X_{t,i_1i_2}^{\Lambda}(3,4;1,2)+X_{u,i_1i_2}^{\Lambda}(3,4;1,2)\Big]\notag\\
&-\sum_j\Big[X_{s,i_1j}^{\Lambda}(1',4;1,3)+X_{u,i_1j}^{\Lambda}(1',4;1,3)\Big]\tilde\Gamma_{ji_2}^{\Lambda}(3,2';4,2)
-\sum_j\tilde\Gamma_{i_1j}^{\Lambda}(1',4;1,3)\Big[X_{s,ji_2}^{\Lambda}(3,2';4,2)+X_{u,ji_2}^{\Lambda}(3,2';4,2)\Big]\notag\\
&+\Big[X_{s,i_1i_2}^{\Lambda}(1',4;1,3)+X_{u,i_1i_2}^{\Lambda}(1',4;1,3)\Big]\tilde\Gamma_{i_2i_2}^{\Lambda}(3,2';2,4)+\tilde\Gamma_{i_1i_2}^{\Lambda}(1',4;1,3)\Big[X_{s,i_2i_2}^{\Lambda}(3,2';2,4)+X_{t,i_2i_2}^{\Lambda}(3,2';2,4)\Big]\notag\\
&+\Big[X_{s,i_1i_1}^{\Lambda}(1',4;3,1)+X_{t,i_1i_1}^{\Lambda}(1',4;3,1)\Big]\tilde\Gamma_{i_1i_2}^{\Lambda}(3,2';4,2)+\tilde\Gamma_{i_1i_1}^{\Lambda}(1',4;3,1)\Big[X_{s,i_1i_2}^{\Lambda}(3,2';4,2)+X_{u,i_1i_2}^{\Lambda}(3,2';4,2)\Big]\notag\\
&+\Big[X_{s,i_1i_2}^{\Lambda}(3,2';1,4)+X_{t,i_1i_2}^{\Lambda}(3,2';1,4)\Big]\tilde\Gamma_{i_1i_2}^{\Lambda}(1',4;3,2)+\tilde\Gamma_{i_1i_2}^{\Lambda}(3,2';1,4)\Big[X_{s,i_1i_2}^{\Lambda}(1',4;3,2)+X_{t,i_1i_2}^{\Lambda}(1',4;3,2)\Big]\Big\}\notag\\
&\times G^\Lambda(\omega_3)G^\Lambda(\omega_4)\;.\label{FRG_twoloop}
\end{align} 
\end{widetext}
Here, the index ``tl'' specifies that the equation only shows the contributions to $d\tilde\Gamma^\Lambda/d\Lambda$ from the two-loop extension. Since this equation has the same form as Eq.~(\ref{FRG_gamma}) but with modified vertex functions, it can be considered to be of {\it effective} one-loop structure. Most importantly, it only contains terms that are already known from Eq.~(\ref{FRG_gamma}), which simplifies its numerical evaluation significantly. Hence, this two-loop scheme is an efficient and numerically not too costly approach which takes into account important (but not all) parts of the three-particle vertex $\Gamma_3^\Lambda$ without the need to solve an explicit flow equation for $\Gamma_3^\Lambda$.

The interaction channels $X^\Lambda_{s/t/u}$ which enter Eq.~(\ref{FRG_twoloop}) contain a single scale propagator $S^\Lambda$, see Eqs. (\ref{xs})-(\ref{xu}). Generally, $S^\Lambda$ can either be implemented as in the one-loop approach [see Eq.~(\ref{singlescale})] or as in the Katanin truncation [see Eq.~(\ref{katanin})]. Using the definition (\ref{singlescale}) for the single-scale propagator immediately leads to the two-loop approach of Ref.~\onlinecite{eberlein14} which can be shown to be exact up to the third order in the effective interaction $\Gamma^\Lambda$ (see Appendix ~\ref{app1}). However, since the last term in Eq.~(\ref{katanin}) is already known from the self-energy flow, it does not come with additional numerical costs to build in Katanin corrections in Eq.~(\ref{FRG_twoloop}). In fact, the inclusion of Katanin terms in Eq.~(\ref{FRG_twoloop}) yields proper contributions of the form $\sim\sum_3 \Gamma_3^\Lambda ( 1',2',3;1,2,3) S^{\Lambda}(\omega_3)$, see Fig.~\ref{fig:nested_diagrams}(d) for an example. In similarity to Sec.~\ref{sec:oneloop}, these corrections ensure a fully self-consistent treatment of the self energy which is important for a proper incorporation of fluctuation effects. All results presented below have been obtained within this ``Katanin-corrected'' two-loop scheme. Its explicit derivation starting from the flow equation for the three-particle vertex is outlined in Appendix~\ref{app1}, where we closely follow Ref.~\onlinecite{eberlein14}. The example in Fig.~\ref{fig:nested_diagrams}(d), showing the graph in Fig.~\ref{fig:nested_diagrams}(b) with a self-energy insertion in the single-scale propagator, illustrates that the extra Katanin-corrected two-loop terms are of forth order in the effective interaction $\Gamma^\Lambda$ and of three-loop type.
\begin{figure*}[t]
\includegraphics[width=0.99\linewidth]{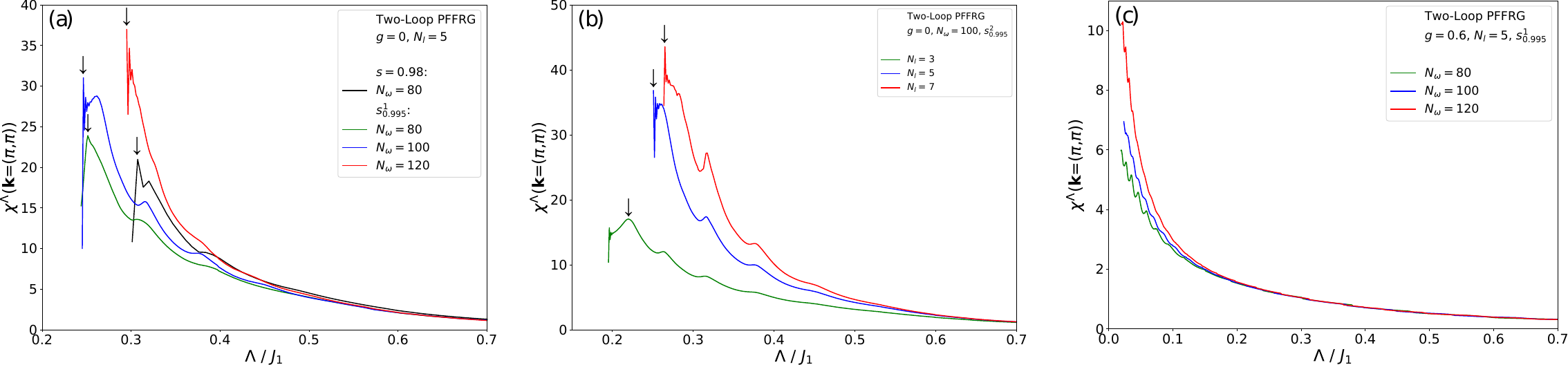}
\caption{Effects of different system sizes $N_\text{l}$, number of frequency mesh points $N_\omega$, and $\Lambda$-integration step widths $s$ on the Katanin-corrected two-loop PFFRG results. (a) N\'eel susceptibility $\chi^\Lambda(\mathbf{k}=(\pi,\pi))$ of the antiferromagnetic nearest neighbor Heisenberg model on the square lattice for varying step widths functions $s$ and frequency meshes $N_\omega$. Arrows indicate the magnetic instability. (b) Dependence of the N\'eel susceptibility on different system sizes $N_\text{l}$. (c) N\'eel susceptibility $\chi^\Lambda(\mathbf{k}=(\pi,\pi))$ of the $J_1$-$J_2$ Heisenberg model at $g=0.6$ for different frequency meshes $N_\omega$. Smooth flows indicate a magnetically disordered phase. \label{fig:check_convergence}}
\end{figure*}

\subsection{Numerical implementation and calculation of magnetic susceptibilities}
\label{sec:numerics}
Before we show results for the two-loop scheme in the next section, some comments about the numerical evaluation of the flow equations and the calculation of physical observables are in order. All approximation schemes discussed in Secs.~\ref{sec:oneloop} and \ref{sec:twoloop} are formulated such that the infinite hierarchy of coupled FRG equations is reduced to a closed set involving only $\Sigma^\Lambda$ and $\tilde\Gamma^\Lambda$. However, to be amenable to numerical treatment further approximations are necessary. Firstly, the FRG equations contain two-particle vertices $\tilde\Gamma^\Lambda_{i_1i_2}$ for all possible combinations of sites $i_1$, $i_2$. A finite set of vertices is obtained by discarding all $\tilde\Gamma^\Lambda_{i_1i_2}$ where the distance between sites $i_1$ and $i_2$ exceeds a given value $N_\text{l}$ (in units of nearest neighbor lattice spacings). This effectively limits the range of spin-spin correlations in the system. Secondly, the Matsubara frequencies $\omega_1$, $\omega_2$, $\omega_{1'}$, $\omega_{2'}$ appearing in the arguments of $\tilde\Gamma^\Lambda_{i_1i_2}(1',2';1,2)$ become continuous at $T=0$. To numerically handle this situation, we define vertex functions on a discrete frequency grid which consists of a finite number of values $N_\omega$ for each of the three transfer frequencies $s$, $t$, and $u$.

Finally, for an actual solution, the RG equations are numerically integrated over $\Lambda$, which requires a small but finite integration step width. Typically, RG steps are defined by a series of values $\ldots,\Lambda_{n-1},\Lambda_{n},\Lambda_{n+1},\ldots$ related via $\Lambda_{n+1}=s\Lambda_{n}$ where $s$ is smaller but close to one. Note that $s$ can be chosen as a constant or it can be a function of the RG scale, $s=s(\Lambda_n)$. The latter possibility allows us to study special points in the RG flow (such as magnetic instabilities) with higher precision, i.e. with smaller integration step widths. Below, we will study different choices for the dependence $s(\Lambda_n)$ which we label by $s(\Lambda_n)=s_x^y(\Lambda_n)$. In this notation $x$ specifies the maximum of the function $s_x^y(\Lambda_n)$ (which occurs in the small $\Lambda$-limit) and $y$ enumerates different functions with the same maximum. The precise form of the functions $s_{0.995}^1$, $s_{0.995}^2$, $s_{0.995}^3$ used below are given in Appendix~\ref{app3}. 

The quality of the results crucially depends on whether a good compromise between the range of correlations $N_\text{l}$, the number of discrete frequencies $N_\omega$ and the RG-integration step width $s$ can be found. For all truncation schemes of Sec.~\ref{sec:method}, the computation times grow with the $2d$-th power in $N_\text{l}$ (where $d$ is the dimension of the system), the fourth power in $N_\omega$ and linearly in the number of $\Lambda$-integration steps. As discussed below, identifying a suitable parameter setting is particularly important in the two-loop scheme. While this approach circumvents the explicit evaluation of a flow equation for the three-particle vertex, this comes at the cost of a rapid error propagation in the derivation of nested graphs. An example for a good parameter choice in the two-loop scheme is $N_\text{l}=5$, $N_\omega=120$, $s=0.995$. Comparing numerical performances for equal parameter settings, the computation times increase roughly by a factor two when including the two-loop contributions.

The physical outcome of the PFFRG approach is the static spin-spin correlator
\begin{equation}
 \chi_{ij}^{zz}=\int_0^{\infty} d\tau \left < S^{z}_i(\tau)S^{z}_j(0) \right >\;,\label{correlator}
\end{equation}
where $\tau$ is an imaginary time variable. Note that for the spin-isotropic models considered here, one has $\chi_{ij}\equiv \chi_{ij}^{xx}= \chi_{ij}^{yy}= \chi_{ij}^{zz}$. Expressing the spin operators in terms of pseudofermions, Eq.~(\ref{correlator}) can be written as a frequency convolution of the two-particle vertex and fermionic propagators. To investigate whether a particular type of magnetic order develops, one calculates the Fourier-transformed correlator $\chi^\Lambda(\mathbf{k})$ (i.e. the momentum dependent magnetic susceptibility) at the corresponding wave vector $\mathbf{k}$. If $\chi^\Lambda(\mathbf{k})$ shows a pronounced kink or cusp during the RG flow this indicates the onset of magnetic order. As argued in Ref.~\onlinecite{iqbal16_3}, the RG scale $\Lambda_\text{c}$ of this cusp may be related to the critical temperature $T_c=\pi \Lambda_\text{c}/2$ at which the magnetic instability occurs. If, on the other hand, the RG flow of the susceptibility remains smooth down to $\Lambda\rightarrow0$ a non-magnetic phase is identified.

\section{Results}\label{sec:results}

\subsection{Antiferromagnetic $J_1$-$J_2$ square lattice Heisenberg model}\label{sec:j1j2}
A main difficulty in the numerical evaluation of the Katanin-corrected two-loop PFFRG equations lies in the rapid error propagation of the nested graph construction. Small errors in the one-loop terms may grow significantly when recombining them into two-loop terms. It is therefore crucial to study how different parameter choices for $N_\text{l}$, $N_\omega$, and $s$ affect the results. To this end, we first consider the antiferromagnetic Heisenberg model on the square lattice with nearest neighbor couplings $J_1>0$ only, see Figs.~\ref{fig:check_convergence}(a) and (b).

For all parameter setting that we have studied, the flow of the $\mathbf{k}=(\pi,\pi)$ component of the magnetic susceptibility $\chi^\Lambda(\mathbf{k})$ shows a pronounced peak during the RG flow [marked by arrows in Figs.~\ref{fig:check_convergence}(a) and (b)] followed by sudden drop and a numerically unstable behavior. This is the expected RG flow behavior in the N\'eel ordered phase, indicating that the magnetic instability is correctly detected. We note that in contrast to an exact solution where susceptibilities should show a real divergence at a magnetic instability, in our PFFRG data we typically see a finite peak. This is because of the combined effects of finite system sizes $N_\text{l}$ and finite frequency grids $N_\omega$ which both regularize the divergence.

The precise shape of the flowing susceptibility shows some characteristic dependencies on $N_\text{l}$, $N_\omega$, and $s$. Most importantly, an insufficient $\Lambda$-integration step width $s$ may lead to an unstable flow behavior before the physically relevant $\Lambda$-regime is reached. An example for $s=0.98$ is shown in Fig.~\ref{fig:check_convergence}(a) where strong and diverging oscillations set in at $\Lambda\approx0.31$ while for a denser integration grid the flow continues to smaller $\Lambda$ values (green line). To obtain reasonable RG flows we find that $s$ needs at least to be given by $0.995$ at small $\Lambda$ (which is fulfilled for the step-width functions $s_{0.995}^{1/2/3}$ used below). In contrast, for the one-loop plus Katanin scheme, an integration step width of $s=0.98$ is often sufficient to yield well-converged results. This stricter condition on $s$ is one of the main reasons for the longer computation times in the two-loop scheme.

We have also checked the influence of different frequency grids, see Fig.~\ref{fig:check_convergence}(a). In general, with increasing number of discrete mesh points the RG flow becomes steeper and the magnetic instability sets in at larger $\Lambda$. Furthermore, for small $N_\omega=80$ or 100 the flowing susceptibility is overlaid by oscillations which, for example, produce the humps at $\Lambda\approx0.32$ in Fig.~\ref{fig:check_convergence}(a). Such features directly reflect the frequency discretization. Our results for denser grids indicate that they mostly disappear for $N_\omega\geq120$. With increasing system sizes $N_\text{l}$ we likewise observe a steeper RG flow and an earlier onset of the instability, see Fig.~\ref{fig:check_convergence}(b). For $N_\text{l}\approx7$, the critical $\Lambda$-scale seems to be mostly converged.
\begin{figure*}[t]
\includegraphics[width=0.99\linewidth]{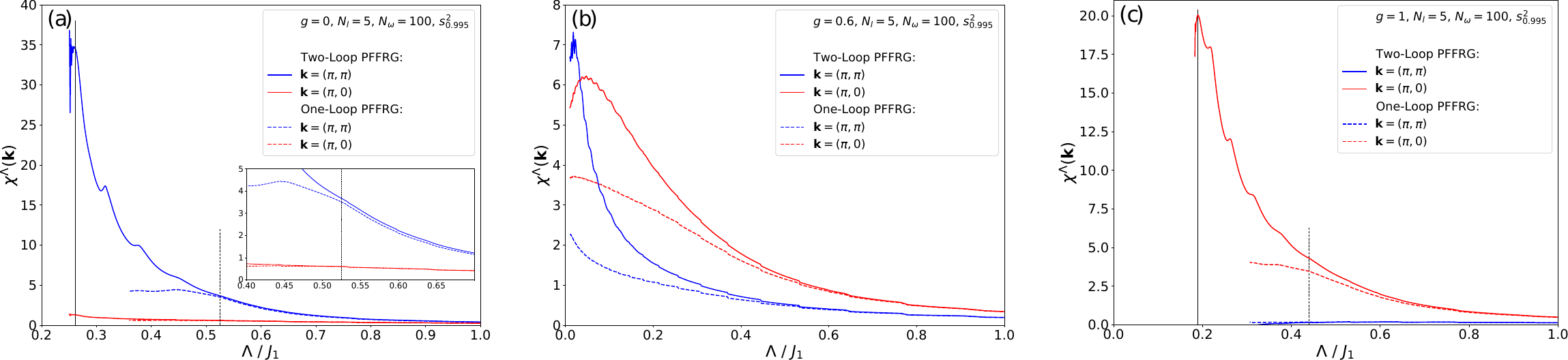}
\caption{Comparison of the flowing susceptibilities in the one-loop plus Katanin (dashed lines) and in the Katanin-corrected two-loop scheme (full lines). Here and in the following figures, the two approaches are briefly denoted by ''one-loop`` and ``two-loop''. Blue (red) lines show the N\'eel (collinear) susceptibility. The plots in (a) - (c) correspond to $g=0$, $g=0.6$, and $g=1$, respectively. Vertical black lines (full and dashed) in (a) and (c) mark the positions of critical RG scales $\Lambda_\text{c}$ in both approaches. The inset in (a) shows an enlarged view of the one-loop instability.\label{fig:compare_old_new}}
\end{figure*}

We repeated this analysis for finite second neighbor interactions $J_2$. In agreement with previous one-loop PFFRG studies, for sufficiently large $g=J_2/J_1$ we identify a non-magnetic phase where the RG flow does not show any instability features down to smallest $\Lambda$-values. An example is given in Fig.~\ref{fig:check_convergence}(c) showing flowing susceptibilities for $g=0.6$ at the N\'eel wave vector $\mathbf{k}=(\pi,\pi)$ (this is the Fourier-component where the susceptibility is maximal). Due to small correlations lengths in magnetically disordered phases the conditions on the system size $N_\text{l}$ are less strict such that $N_\text{l}=5$ is mostly sufficient. Also the convergence in $N_\omega$ is found to be better. Nevertheless, at very small $\Lambda\lesssim0.1$ good convergence is hard to reach even for $N_\omega=120$.

Further insight is gained when comparing the RG flows of the Katanin-corrected two-loop scheme with the previous one-loop plus Katanin approach. Differences are most obvious in magnetically ordered phases such as for $g=0$ [Fig.~\ref{fig:compare_old_new}(a)] and $g=1$  [Fig.~\ref{fig:compare_old_new}(c)]. In the latter case, an instability at $\mathbf{k}=(\pi,0)$ clearly indicates the collinear phase. Considering $\chi^\Lambda(\mathbf{k}=(\pi,\pi))$ for $g=0$ and $\chi^\Lambda(\mathbf{k}=(\pi,0))$ for $g=1$ (at other wave vectors the susceptibilities are comparatively small) we find that down to the critical RG scale $\Lambda_\text{c}$ of the one-loop PFFRG scheme, both susceptibilities are of very similar size. Below this point, the two-loop result keeps increasing until an instability occurs at a critical $\Lambda_\text{c}$ that is roughly halved compared to the one-loop result. Furthermore, instability features appear more sharply. The significant decrease of $\Lambda_\text{c}$ reveals an important thermodynamic property of the two-loop scheme. We first note that both the RG scale $\Lambda$ and the temperature $T$ effectively act as an infrared frequency cutoff. Based on a comparison of the cutoff procedures in the large $S$ limit, it has been argued\cite{iqbal16_3} that they are related via $T=\pi \Lambda/2$. Since the Mermin-Wagner theorem forbids finite-temperature phase transitions in 2D Heisenberg models\cite{mermin66,hohenberg67}, one would not expect to find instabilities at finite RG cutoffs. Critical scales $\Lambda_\text{c}>0$ must, hence, be considered as artifacts of the truncation of FRG equations. The decrease of $\Lambda_\text{c}$, however, indicates that the inclusion of two-loop terms improves the fulfillment of the Mermin-Wagner theorem significantly. Indeed, the origin of this improvement can be explained on a diagrammatic level, as detailed in Appendix~\ref{app2}.

We also compare one-loop and two-loop results in the magnetically disordered phase at $g=0.6$, see Fig.~\ref{fig:compare_old_new}(b). While down to $\Lambda\approx0.4$ both approaches yield similar results, for smaller $\Lambda$ the two-loop susceptibility becomes larger. However, within wide ranges of $\Lambda$, the ratio of the susceptibility at wave vectors $\mathbf{k}=(\pi,\pi)$ and $\mathbf{k}=(\pi,0)$ remains roughly unchanged when including two-loop contributions. 
\begin{figure}[t]
\includegraphics[width=0.8\linewidth]{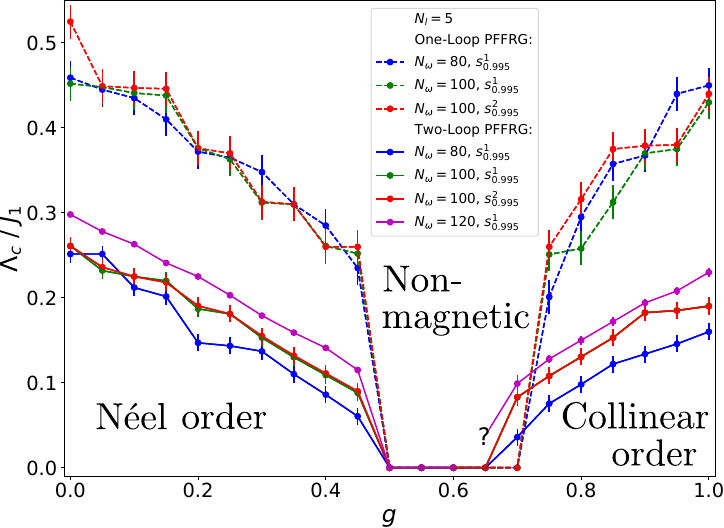}
\caption{Phase diagram of the $J_1$-$J_2$ square lattice Heisenberg model showing the dependence of the critical RG scale $\Lambda_\text{c}$ on the coupling ratio $g$. A vanishing $\Lambda_\text{c}$ indicates a magnetically disordered phase. Shown are results for the one-loop (dashed lines) and two-loop (full lines) approaches using different frequency grids $N_\text{l}$ and RG step sizes. A data point with large numerical uncertainties is marked by ``?''.\label{fig:phase_diag}}
\end{figure}

We have investigated more coupling ratios $g$ to map out the phase diagram in the range $0\leq g\leq1$, see Fig.~\ref{fig:phase_diag}. We find that the overall sequence of quantum phases and their boundaries remain largely unchanged when varying the $\Lambda$-integration step widths and the frequency meshes. This also applies to differences between the one-loop and two-loop schemes. Furthermore, the aforementioned factor of $\sim2$ between the critical scales of both approaches appears throughout the ordered phases.

It is important to emphasize that an exact determination of phase boundaries is a difficult task in all PFFRG schemes. This is because close to critical points where instability scales $\Lambda_\text{c}$ become small, it is hard to distinguish whether an observed anomaly is due to numerical errors (which inevitably grow at small $\Lambda$) or due to a real magnetic instability. For this reason we have not attempted to investigate the critical regions with higher precision. Some quantitative conclusions about the positions of the phase boundaries can still be drawn. Firstly, we do not find any shifts of the transition at $g_{\text{c}1}$ when including two-loop terms. A rough estimate of its position yields $0.45<g_{\text{c}1}<0.5$. Slight differences compared to the results reported in an earlier study\cite{reuther10} are due to the denser frequency and $\Lambda$ grids used here. On the other hand, for the second transition at $g_{\text{c}2}$ our results indicate small modifications. While the one-loop results would be consistent with a transition at around $g_{\text{c}2}\approx0.7$, the two-loop scheme still detects a clear magnetic instability feature at this point. A calculation with an increased number of frequencies ($N_\omega=120$) might even point towards collinear magnetic order at $g=0.65$, although uncertainties are significant due to the aforementioned reason (this data point is marked by ``?'' in Fig.~\ref{fig:phase_diag}). We, hence, estimate this phase transition to be approximately located at $g_{\text{c}2}\approx0.65$ within the two-loop approach.

Comparing with the literature, other numerical works have reported a large variety of different boundary positions\cite{jiang12,schulz96,isaev09,darradi08,gong14,richter10,poilblanc17,wang17,wang16,hu13,singh99,kotov99} which do not yet allow to draw a final conclusion about the exact extent of the ground state phases. While for the first transition, earlier studies have favored $g_{\text{c}1}\approx0.4$ or even smaller\cite{schulz96,singh99,kotov99}, more recent works find larger values such as $g_{\text{c}1}\approx0.45$\cite{gong14,wang17}. Very recent PEPS approaches\cite{poilblanc17} predict the transition to be close to $0.5$ or even above\cite{wang16,haghshenas17}. Therefore, the relatively large value of $0.45<g_{\text{c}1}<0.5$ that we find seems to be in good agreement with the latest numerical studies. Concerning the second phase transition, our estimate of $g_{\text{c}2}\approx0.65$ is certainly in the upper range of values predicted by other methods but is still consistent with Refs.~\onlinecite{schulz96,isaev09,richter10}. Particularly, the reduction of $g_{\text{c}2}$ upon including two-loop contributions improves the agreement with other works. 
\begin{figure}[t]
\includegraphics[width=0.8\linewidth]{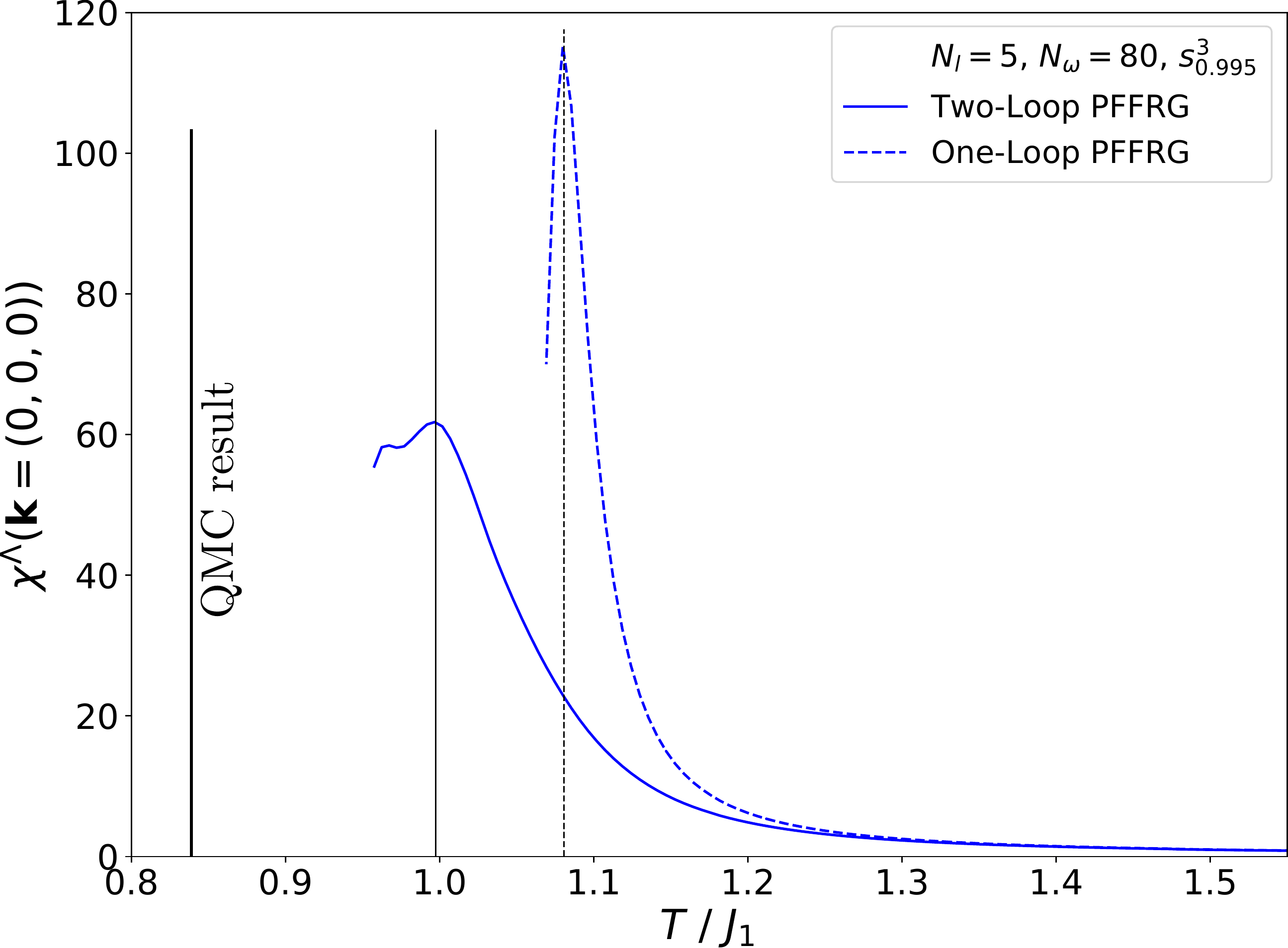}
\caption{Susceptibility $\chi^\Lambda(\mathbf{k}=(0,0,0))$ of the nearest neighbor ferromagnetic Heisenberg model on the simple cubic lattice. Dashed (full) lines denote the one-loop (two-loop) results. Note that the $x$-axis has been converted into a temperature axis by using the relation $T=\pi \Lambda/2$. Vertical lines (dashed and full) mark the critical temperatures of the one-loop and two-loop approaches, which are given by $T_\text{c}=1.081|J_1|$ and $T_\text{c}=0.997|J_1|$, respectively. The thick black line denotes the exact quantum Monte Carlo result\cite{troyer04} $T_\text{c}=0.839|J_1|$.\label{fig:sc_fm}}
\end{figure}

\subsection{Ferromagnetic Heisenberg model on the simple cubic lattice}\label{sec:sc_fm}
We conclude this section with a brief discussion of critical temperatures in the 3D case where the Mermin-Wagner theorem does not forbid magnetic instabilities at finite $T$. As an example, we consider the nearest neighbor ferromagnetic Heisenberg model on the simple cubic lattice. The non-frustrated nature of this model allows us to compare results with the (quasi-) exact quantum Monte-Carlo method. Apart from increased numerical efforts due to additional site summations (which lead to a scaling of computation times with the sixth power in $N_\text{l}$ instead of the fourth power) the PFFRG schemes do not undergo any conceptual modifications in 3D. In Fig.~\ref{fig:sc_fm} we show the flowing susceptibility $\chi^\Lambda(\mathbf{k}=(0,0,0))$ for the one-loop and two-loop schemes. While both approaches clearly detect a ferromagnetic instability during the RG flows, the critical RG scale of the two-loop scheme is found to be smaller. The relative reduction of $\Lambda_\text{c}$, however, is not as pronounced as in the 2D case. Furthermore, in contrast to 2D where both schemes only differ below the one-loop critical RG scale, here two-loop contributions seem to become relevant much earlier. For comparison with the quantum Monte Carlo result\cite{troyer04} $T_\text{c}=0.839|J_1|$, we use the relation $T_c=\pi \Lambda_\text{c}/2$ that has been proposed in Ref.~\onlinecite{iqbal16_3}. We find that the one-loop critical temperature $T_\text{c}=1.081|J_1|$ overestimates the quantum Monte Carlo value by $\sim29\%$ while the two-loop result $T_\text{c}=0.997|J_1|$ only overestimates the exact value by $\sim19\%$. In similarity to the 2D case this indicates that critical temperatures come out significantly better when including two-loop terms.

\section{Conclusion and outlook}\label{sec:conclusion}
In this work we have developed an extension of the PFFRG approach that takes into account two-loop diagrammatic contributions. Conceptually, the additional terms are generated by recombining two-particle interaction channels such that the approach has an effective one-loop structure. To ensure a self-consistent treatment of self-energy effects our approach also includes Katanin-corrections of three-loop type. Compared to a PFFRG scheme that explicitly takes into account the flow of three-particle vertex functions (which has not yet been implemented), this formulation simplifies the numerical evaluation significantly. Yet, our two-loop scheme suffers from severe error propagation which we mitigate by using dense frequency meshes and small RG integration step widths.

As a first exploratory study to benchmark our results, we consider the antiferromagnetic $J_1$-$J_2$ Heisenberg model on the square lattice. Despite the fact that the two-loop extension involves a large number of new terms in the RG equations, the overall phase diagram of this model remains surprisingly unchanged. Particularly, we clearly find the expected sequence of N\'eel ordered, non-magnetic and collinear ordered phases. While we do not detect any changes of the transition from the N\'eel ordered to the non-magnetic phase upon including two-loop terms, the boundary between the non-magnetic and the collinear phase undergoes a moderate shift towards smaller coupling ratios $g$. This result indicates that the one-loop approach already yields good approximations of magnetic phase diagrams. In contrast, critical RG scales $\Lambda_\text{c}$ (which are finite in all PFFRG schemes) change significantly in our two-loop extension. Throughout the ordered phases we find that $\Lambda_\text{c}$ (or equivalently the critical temperature $T_\text{c}$) is reduced by a factor $\sim2$ which indicates a better fulfillment of the Mermin-Wagner theorem. We explain this improvement by identifying the relevant self energy and two-particle vertex diagrams. Better estimates for critical temperatures are also obtained in 3D such as for the ferromagnetic nearest neighbor Heisenberg model on the simple cubic lattice. While the one-loop and two-loop schemes both overestimate $T_\text{c}$, the error is reduced by $\sim1/3$ in the two-loop approach.

Our two-loop extension will be useful for a variety of future investigations. For example, it would be interesting to study the nature of the non-magnetic phase in the $J_1$-$J_2$ Heisenberg model which has not been further characterized here. A previous one-loop PFFRG work\cite{reuther10} investigated dimer susceptibilities for various valence-bond crystal configurations and found that the responses are generally rather small. Furthermore, the responses for different dimer configurations are almost identical which would be consistent with a spin-liquid ground state. Whether a certain valence-bond crystal is favored by additional two-loop contributions remains an open question for future studies. Given the large number of interesting frustrated spin systems in 2D and 3D we also suggest to apply our approach to further lattice models to check whether the conclusions of this work remain valid. Examples for possible lattices include the triangular, honeycomb and kagome lattices. The ultimate goal of two-loop PFFRG schemes would be to describe chiral magnetic properties such as the formation of chiral spin liquids. Since the spin-chirality term involves three spin operators, this is not possible in a pure two-particle (or one-loop) formalism. Whether the current two-loop extension is sufficient to resolve such effects is {\it a priori} not clear and needs to be investigated in detail.
 
\section{Acknowledgements}
We thank Max Hering, Yasir Iqbal, Ronny Thomale, and Simon Trebst for fruitful discussions. This work is supported by the Freie Universit\"at Berlin within the Excellence Initiative of the German Research Foundation.
\appendix
\section{Derivation of the two-loop truncation from the three-particle vertex flow}
\label{app1}
 \begin{figure}[t]
\includegraphics[width=0.99\linewidth]{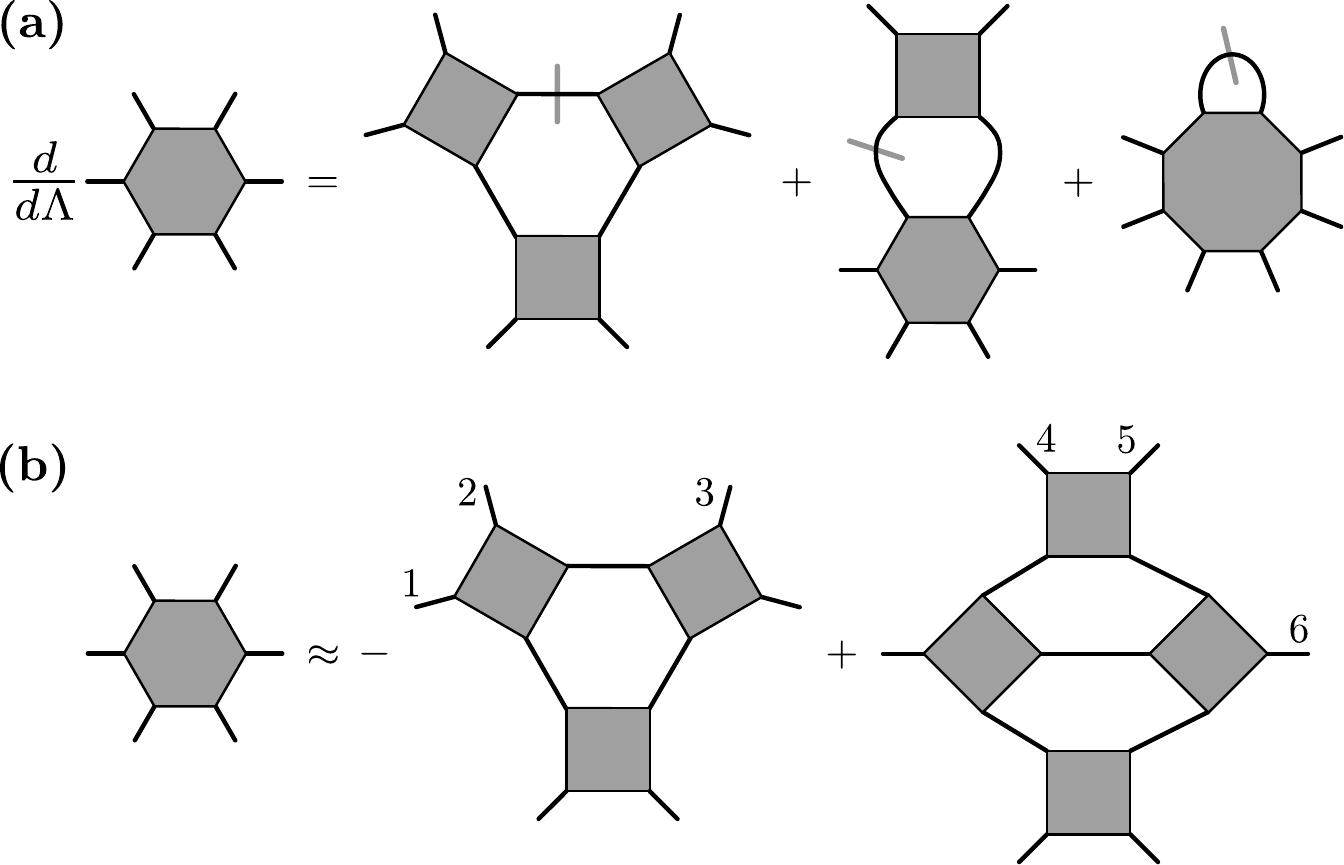}
\caption{(a) Schematic diagrammatic illustration of the FRG equation for the three-particle vertex. For simplicity we draw the fermion lines without arrows and do not specify how fermion lines are connected inside a vertex. Note that in the first and second term on the right hand side, the single-scale propagators may also appear at another propagator line. (b) Approximation of the three-particle vertex as derived in Eq.~(\ref{approx_three_part}). All truncation schemes of Sec.~\ref{sec:method} are based on the depicted diagrams, see text for details. \label{fig:three_part_flow}}
\end{figure}
In this appendix we outline how the two-loop scheme of Eq.~(\ref{FRG_twoloop}) can be formally derived from the flow equation of the three-particle vertex. We closely follow Ref.~\onlinecite{eberlein14} but also show how the Katanin-corrected two-loop terms of Fig.~\ref{fig:nested_diagrams}(d) are obtained.

We start with the flow equation for the three-particle vertex which can be schematically written as
\begin{align}
\frac{d}{d\Lambda}\Gamma_3^\Lambda&=\text{tr}\left(S^\Lambda \Gamma^\Lambda G^\Lambda \Gamma^\Lambda G^\Lambda \Gamma^\Lambda+G^\Lambda \Gamma^\Lambda S^\Lambda \Gamma^\Lambda G^\Lambda \Gamma^\Lambda\right.\notag\\
&+\left.G^\Lambda \Gamma^\Lambda G^\Lambda \Gamma^\Lambda S^\Lambda \Gamma^\Lambda\right)\notag\\
&+\text{tr}\left[\Gamma^\Lambda\left(S^\Lambda G^\Lambda+G^\Lambda S^\Lambda\right)\Gamma_3^\Lambda\right]\notag\\
&+\text{tr}\left(S^\Lambda \Gamma_4^\Lambda \right)\,.\label{FRG_threepart}
\end{align}
For simplicity, we have omitted all arguments of the vertex functions. The trace stands for the internal summations/integrations and $\Gamma_4^\Lambda$ is the four-particle vertex. A diagrammatic illustration of this equation is shown in Fig.~\ref{fig:three_part_flow}. The single-scale propagator in Eq.~(\ref{FRG_threepart}) is given by Eq.~(\ref{singlescale}) where the $\Lambda$-derivative only acts on the $\theta$-function contained in $G_0^\Lambda$. One can therefore write
\begin{equation}
S^\Lambda=-\tilde G^\Lambda d_\Lambda \theta^\Lambda\quad\text{with}\quad\tilde G^\Lambda=\left(G_0^{-1}-\Sigma^\Lambda\right)^{-1}\;,
\end{equation}
where $\theta^\Lambda\equiv\theta(|\omega|-\Lambda)$ and $d_\Lambda\equiv d/d\Lambda$ are shorthand notations for the regulator function and the $\Lambda$-derivative, respectively. Using this identity, Eq.~(\ref{FRG_threepart}) becomes
\begin{align}
d_\Lambda\Gamma_3^\Lambda&=-\text{tr}\left[\tilde G^\Lambda \Gamma^\Lambda \tilde G^\Lambda \Gamma^\Lambda \tilde G^\Lambda \Gamma^\Lambda \left(d_\Lambda\theta^\Lambda\theta^\Lambda\theta^\Lambda\right)\right]\notag\\
&-\text{tr}\left[\Gamma^\Lambda \tilde G^\Lambda \tilde G^\Lambda\Gamma_3^\Lambda\left(d_\Lambda\theta^\Lambda\theta^\Lambda\right)\right]+\text{tr}\left(S^\Lambda \Gamma_4^\Lambda \right)\,,\label{FRG_threepart2}
\end{align}
where we avoid writing repeating factors as powers to indicate that they might have different arguments. In the following steps of approximation, we only neglect terms in fourth or higher orders in the effective interaction $\Gamma^\Lambda$ such that the scheme remains exact up to $\mathcal{O}((\Gamma^\Lambda)^3)$. Firstly, this allows us to discard the term $\sim\Gamma_4^\Lambda$ in Eq.~(\ref{FRG_threepart2}) since the four-particle vertex is at least of the order $(\Gamma^\Lambda)^4$. As the three-particle vertex is $\sim\mathcal{O}((\Gamma^\Lambda)^3)$, one could also neglect the term $\sim \Gamma^\Lambda\Gamma_3^\Lambda$ within this level of approximation (as done in Ref.~\onlinecite{eberlein14}). However, since this term will generate the Katanin-corrected two-loop diagrams of the order $(\Gamma^\Lambda)^4$, we will keep it here. The next step amounts to rewriting Eq.~(\ref{FRG_threepart2}) such that it contains $\Lambda$-derivatives acting on the entire trace,
\begin{widetext}
\begin{align}
d_\Lambda\Gamma_3^\Lambda&=-d_\Lambda\text{tr}\left(\tilde G^\Lambda \Gamma^\Lambda \tilde G^\Lambda \Gamma^\Lambda \tilde G^\Lambda \Gamma^\Lambda \theta^\Lambda\theta^\Lambda\theta^\Lambda\right)-d_\Lambda\text{tr}\left(\Gamma^\Lambda \tilde G^\Lambda \tilde G^\Lambda\Gamma_3^\Lambda\theta^\Lambda\theta^\Lambda\right)\notag\\
&+\text{tr}\left[\left(d_\Lambda\Gamma^\Lambda\Gamma^\Lambda\Gamma^\Lambda\right)\tilde G^\Lambda \tilde G^\Lambda \tilde G^\Lambda\theta^\Lambda\theta^\Lambda\theta^\Lambda \right]
+\text{tr}\left[\Gamma^\Lambda\Gamma^\Lambda\Gamma^\Lambda\left(d_\Lambda\tilde G^\Lambda \tilde G^\Lambda \tilde G^\Lambda\right)\theta^\Lambda\theta^\Lambda\theta^\Lambda \right]\notag\\
&+\text{tr}\left[\left(d_\Lambda \Gamma^\Lambda\right)\tilde G^\Lambda \tilde G^\Lambda\Gamma_3^\Lambda \theta^\Lambda\theta^\Lambda\right]
+\text{tr}\left[\Gamma^\Lambda\left(d_\Lambda \tilde G^\Lambda \tilde G^\Lambda\right)\Gamma_3^\Lambda \theta^\Lambda\theta^\Lambda\right]
+\text{tr}\left[\Gamma^\Lambda \tilde G^\Lambda \tilde G^\Lambda\left(d_\Lambda \Gamma_3^\Lambda\right)\theta^\Lambda\theta^\Lambda\right]
+\mathcal{O}\left((\Gamma^
\Lambda)^4\right)\,.\label{FRG_threepart3}
\end{align}
\end{widetext}
Using $\tilde G\sim \mathcal{O}((\Gamma^\Lambda)^0)$ and counting the powers of $\Gamma^\Lambda$ appearing on the right hand sides of the flow equations for $\Sigma^\Lambda$, $\Gamma^\Lambda$, and $\Gamma_3^\Lambda$ one finds
\begin{align}
d_\Lambda\tilde G^\Lambda &\sim \left(\tilde G^\Lambda\right)^2 d_\Lambda\Sigma^\Lambda\sim\mathcal{O}(\Gamma^\Lambda)\,,\notag\\
d_\Lambda\Gamma^\Lambda&\sim \mathcal{O}\left((\Gamma^\Lambda)^2\right)\;,\quad d_\Lambda\Gamma^\Lambda_3\sim \mathcal{O}\left((\Gamma^\Lambda)^3\right)\,.
\end{align}
It follows that all terms in the second and third lines of Eq.~(\ref{FRG_threepart3}) are at least on the order $(\Gamma^\Lambda)^4$ and can therefore be neglected within the current level of approximation. Exploiting $G^\Lambda=\tilde G^\Lambda\theta^\Lambda$ one may write Eq.~(\ref{FRG_threepart3}) as
\begin{align}
d_\Lambda\Gamma_3^\Lambda&=-d_\Lambda\text{tr}\left(G^\Lambda \Gamma^\Lambda G^\Lambda \Gamma^\Lambda G^\Lambda \Gamma^\Lambda\right)-d_\Lambda\text{tr}\left(\Gamma^\Lambda G^\Lambda G^\Lambda \Gamma_3^\Lambda\right)\notag\\
&+\mathcal{O}\left((\Gamma^\Lambda)^4\right)\;.
\end{align}
A straightforward $\Lambda$ integration yields a self-consistent equation for $\Gamma_3^\Lambda$,
\begin{equation}
\Gamma_3^\Lambda\approx-\text{tr}\left(G^\Lambda \Gamma^\Lambda G^\Lambda \Gamma^\Lambda G^\Lambda \Gamma^\Lambda\right)-\text{tr}\left(\Gamma^\Lambda G^\Lambda G^\Lambda \Gamma_3^\Lambda\right)\,.\label{sc_three_part}
\end{equation}
This equation may be solved interatively by successively inserting it into itself. The first iteration step leads to
\begin{align}
\Gamma_3^\Lambda&\approx-\text{tr}\left(G^\Lambda \Gamma^\Lambda G^\Lambda \Gamma^\Lambda G^\Lambda \Gamma^\Lambda\right)\notag\\
&+\text{tr}\left[\Gamma^\Lambda G^\Lambda G^\Lambda   \text{tr}\left(G^\Lambda \Gamma^\Lambda G^\Lambda \Gamma^\Lambda G^\Lambda \Gamma^\Lambda\right)  \right]\,.\label{approx_three_part}
\end{align}
All truncation schemes discussed in Sec.~\ref{sec:method} (i.e., the Katanin truncation of Sec.~\ref{sec:oneloop}, the two-loop extension of Sec.~\ref{sec:twoloop} and the Katanin-corrected two-loop scheme) are based on this approximation for $\Gamma_3^\Lambda$. For a diagrammatic representation of Eq.~(\ref{approx_three_part}), see Fig.~\ref{fig:three_part_flow}(b). To obtain a contribution to $\sim\sum_3 \Gamma_3^\Lambda ( 1',2',3;1,2,3) S^{\Lambda}(\omega_3)$ in Eq.~(\ref{FRG_gamma}), a pair of external propagator lines of $\Gamma_3^\Lambda$ needs to be connected by a single-scale propagator. As illustrated in Fig.~\ref{fig:three_part_flow}(b) there are various ways of performing such contractions. Considering the first term $\sim\Gamma^\Lambda\Gamma^\Lambda\Gamma^\Lambda$, one may either fuse two lines belonging to the same two-particle vertex [such as lines ``1'' and ``2'' in Fig.~\ref{fig:three_part_flow}(b)] or two lines belonging to different two-particle vertices [such as lines ``2'' and ``3'' in Fig.~\ref{fig:three_part_flow}(b)]. While the first possibility leads to a Katanin term as in Fig.~\ref{fig:nested_diagrams}(a), the second possibility generates a nested two-loop diagram as in Fig.~\ref{fig:nested_diagrams}(b). Similar types of contractions also exist for the second term in Eq.~(\ref{approx_three_part}) which already contributes to the fourth order in $\Gamma^\Lambda$. Connecting lines ``4'' and ``5'' in Fig.~\ref{fig:three_part_flow}(b) yields a Katanin-corrected two-loop diagram such as the one in Fig.~\ref{fig:nested_diagrams}(d). However, other possibilities of connecting fermion lines, e.g., ``4'' and ``6'' are not taken into account since such diagrams cannot be simply expressed as a recombination of two interaction channels.

\section{Diagrammatic discussion of the Mermin-Wagner theorem in one and two-loop PFFRG schemes}
\label{app2}
 \begin{figure*}[t]
\includegraphics[width=0.9\linewidth]{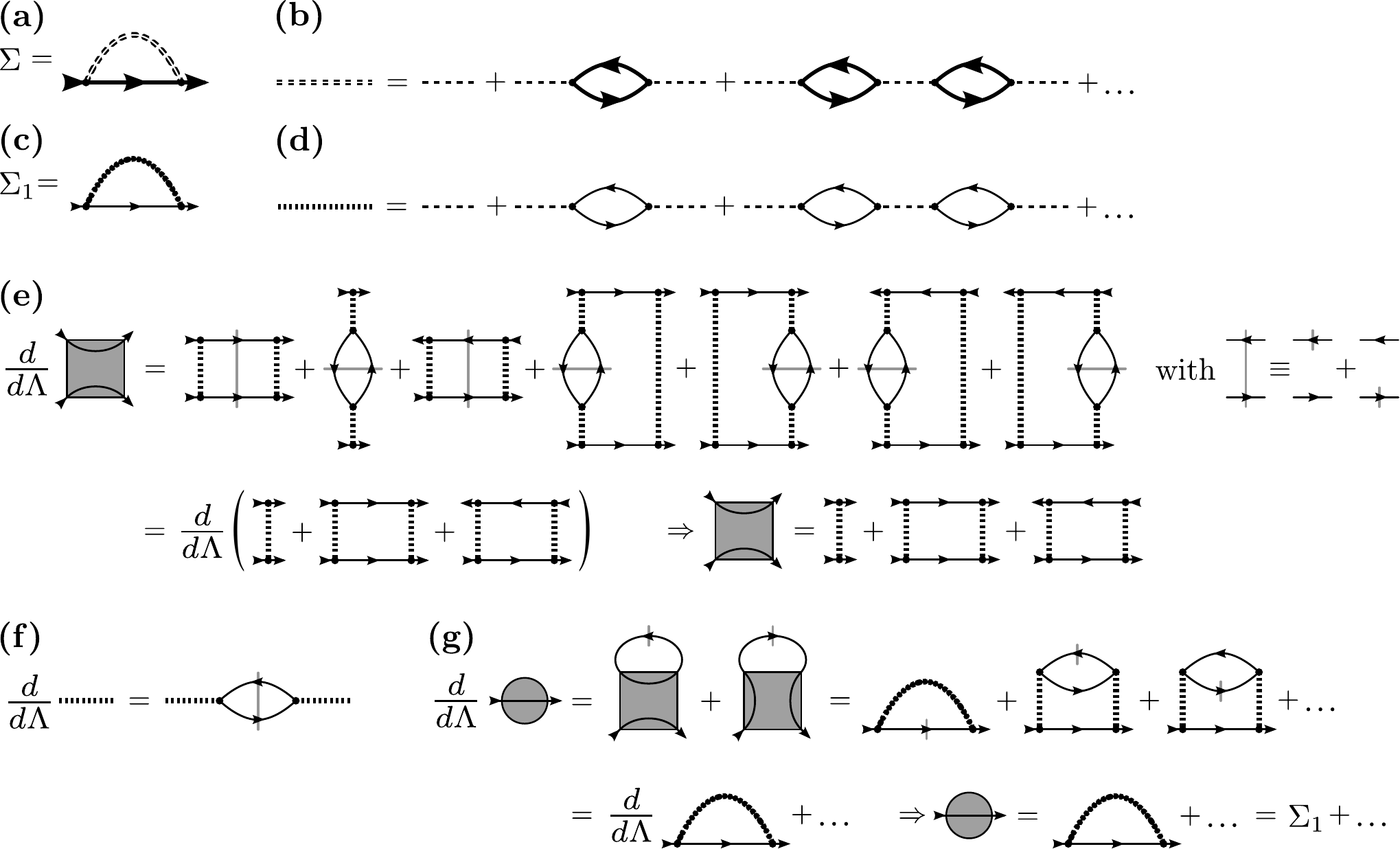}
\caption{(a) Self energy of the BW approximation. The thick line with an arrow is the dressed propagator in Eq.~(\ref{ds}). The double dashed line is a renormalized interaction defined by the RPA series in (b). The thin dashed line in (b) is the bare interaction $J_{ij}$. As explained in the main text, the self-consistent structure of (a) and (b) guarantees the fulfillment of the Mermin-Wagner theorem. (c) On the first level of iteration, the self-energy $\Sigma_1$ carries a free fermion propagator (thin line with an arrow) and a ``partially'' renormalized interaction (thick dashed line), referred to as bare RPA interaction. (d) The bare RPA interaction follows from an RPA series containing free fermion propagators. (e) Simplified version of the two-loop PFFRG flow equation where on the right hand side the two-particle vertex has been replaced by the bare RPA interaction and only certain interaction channels are taken into account. This equation can be integrated, leading to the two-particle diagrams in the second line. (f) $\Lambda$-derivative of the bare RPA interaction. (g) PFFRG flow equation for the self energy with the two-particle diagrams from (e) inserted (only three out of six terms are shown in the first line). The equation can be integrated proving that $\Sigma_1$ is fully contained in a two-loop PFFRG scheme.\label{fig:mermin_wagner}}
\end{figure*}
In this appendix we present a diagrammatic argument explaining why the two-loop scheme of Sec.~\ref{sec:twoloop} leads to an improved fulfillment of the Mermin-Wagner theorem (i.e., to reduced critical temperatures in isotropic 2D Heisenberg systems). Our argument relies on an approximation for the two-particle vertex which has been proposed by Brinckmann and W\"olfle\cite{brinckmann04} (we call it the BW approximation in the following). This scheme is similar to the FLEX approximation that has been used to investigate 2D Hubbard models\cite{bickers89}. Most importantly, Brinckmann and W\"olfle showed analytically and numerically that their approximation exactly fulfills the Mermin-Wagner theorem in 2D.

The BW scheme is based on the same pseudofermionic representation as our PFFRG approach. It starts defining a dressed propagator $G(\omega)$ via the Dyson-Schwinger equation
\begin{equation}
G(\omega)=[i\omega-\Sigma(\omega)]^{-1}\,.\label{ds}
\end{equation}
The self energy $\Sigma(\omega)$ that enters Eq.~(\ref{ds}) is of Fock type as depicted in Fig.~\ref{fig:mermin_wagner}(a). Note that the double dashed line in Fig.~\ref{fig:mermin_wagner}(a) is a renormalized exchange interaction generated by the RPA series in Fig.~\ref{fig:mermin_wagner}(b) [the thin dashed line on the right hand side of Fig.~\ref{fig:mermin_wagner}(b) is the bare exchange interaction]. Most importantly, all propagators appearing in Figs.~\ref{fig:mermin_wagner}(a) and (b) are the dressed ones of Eq.~(\ref{ds}), drawn as thick lines. This property makes the BW approximation fully self-consistent: The dressed propagator depends on the self-energy which in turn contains the full propagator. The self-consistency is also the key reason why the BW scheme fulfills the Mermin-Wagner theorem. Within a pseudofermionic formulation, the RPA diagrams represent the leading order in a $1/S$ expansion\cite{baez17}, i.e., they show the strongest divergence when approaching a magnetic instability from the high temperature side. To discuss the Mermin-Wagner theorem in the context of the BW scheme, we first assume that a magnetic instability occurs at a finite temperature $T_\text{c}>0$. When approaching $T_\text{c}$ from above, the diverging effective interaction in Fig.~\ref{fig:mermin_wagner}(b) is fed back into the self energy which, hence, becomes large near criticality. The self energy is purely imaginary due to particle-hole symmetry and can therefore be interpreted as an inverse pseudofermion lifetime. Since a small fermion lifetime is equivalent to strong quantum fluctuations, the instability is suppressed when reinserting the self energy into the RPA diagram (via dressed propagators). In 2D this negative feedback mechanism reduces critical temperatures down to zero in agreement with the Mermin-Wagner theorem\cite{brinckmann04}.

In principle, the BW approximation can be treated iteratively by successively calculating self energies and reinserting them into the RPA diagram. In the following, we will discuss to which extent the diagrams of such iterations are included in our PFFRG approaches. To avoid lengthy formulas and to be more illustrative this analysis is done diagrammatically. We particularly focus on the diagrams of the first iteration step depicted in Figs.~\ref{fig:mermin_wagner}(c) and (d). On this level of approximation, the self energy $\Sigma_1$ contains a free fermion propagator $G_0=1/(i\omega)$ which is drawn as a thin line. Furthermore, $\Sigma_1$ contains a ``partially'' renormalized interaction defined in Fig.~\ref{fig:mermin_wagner}(d) and illustrated by a thick dashed line. This interaction follows from an RPA series that contains the free propagator and the bare interaction (we therefore refer to it as the ``bare RPA interaction'' in the following). Below we will show that $\Sigma_1$ is completely contained in the two-loop PFFRG scheme of Sec.~\ref{sec:twoloop} such that the feedback mechanism suppressing $T_\text{c}$ is at least included on the level of the first iteration step. On the other hand, within the one-loop (plus Katanin) truncation not even $\Sigma_1$ is fully summed up, which explains the significant improvement when including two-loop terms.

Our diagrammatic argument proving that $\Sigma_1$ is fully contained in the extended scheme of Sec.~\ref{sec:twoloop} starts with the two-loop flow equations in Eqs.~(\ref{FRG_gamma}) and (\ref{FRG_twoloop}). We approximate their right hand sides such that they generate a smaller set of diagrams. Note that for all truncation schemes presented in Sec.~\ref{sec:method}, the bare RPA interaction of Fig.~\ref{fig:mermin_wagner}(d) is fully contained in an PFFRG solution for $\tilde\Gamma^\Lambda$. This is proven in Ref.~\onlinecite{baez17}, where it is shown that such diagrams are generated by the RPA interaction channel. In a first step of approximation, all two-particle vertices $\tilde\Gamma^\Lambda$ appearing on the right hand sides of Eqs.~(\ref{FRG_gamma}) and (\ref{xs})-(\ref{FRG_twoloop}) are replaced by the bare RPA interaction. With this replacement, the amount of diagrams contained in a solution for $\tilde\Gamma^\Lambda$ is reduced. In a further reduction, only certain interaction channels are taken into account. Particularly, from the five interaction channels of Eq.~(\ref{FRG_gamma}) only the particle-particle, the RPA, and the particle-hole channels are kept. Furthermore, from the many possibilities of constructing nested two-loop terms only four contributions are considered. Altogether, these modifications result in the two-loop equation for $\tilde\Gamma^\Lambda$ shown in the first line of Fig.~\ref{fig:mermin_wagner}(e). As usual, gray slashes indicate that a $\Lambda$-derivative is acting on the propagator. Most importantly, the product rule for derivatives may be applied backwards on the right hand side of this equation such that the whole expression can be written as a single $\Lambda$-derivative, see second line of Fig.~\ref{fig:mermin_wagner}(e). This step of manipulation uses the identity of Fig.~\ref{fig:mermin_wagner}(f) showing how the $\Lambda$-derivative acts on the bare RPA interaction. The equation in Fig.~\ref{fig:mermin_wagner}(e) may now be straightforwardly $\Lambda$-integrated as a whole. The solution for $\tilde\Gamma^\Lambda$ is given by the three diagrams in the second line of Fig.~\ref{fig:mermin_wagner}(e). Since our manipulations of the flow equation only reduce the amount of generated terms, this proves that such diagrams are certainly contained in a two-loop PFFRG scheme. 

We now consider the PFFRG equation for the self energy in Eq.~(\ref{FRG_sigma}) [which is also depicted in Fig.~\ref{fig:mermin_wagner}(g)]. Replacing the two-particle vertex on the right hand side of this equation by the solution in Fig.~\ref{fig:mermin_wagner}(e) results in six graphs, three of which are shown in the first line of Fig.~\ref{fig:mermin_wagner}(g). The other terms are irrelevant for this discussion. One finds that these three graphs can be written as a $\Lambda$-derivative of $\Sigma_1$ as depicted in the second line of \ref{fig:mermin_wagner}(g). Integrating this equation, hence, shows that $\Sigma_1$ is fully contained in the self energy of a two-loop PFFRG scheme.

For this derivation the inclusion of the nested two-loop graphs in the first line of Fig.~\ref{fig:mermin_wagner}(e) is crucial. Without these contributions (i.e., in a one-loop scheme) $\Sigma_1$ is no longer completely summed up. Similar arguments can also be formulated for higher levels of iteration $\Sigma_2$, $\Sigma_3,\ldots$. It turns out that already $\Sigma_2$ is not fully contained in any of the truncations discussed here. However, at least more diagrammatic contributions of $\Sigma_2$ are summed up when extending the PFFRG from one-loop to two-loop. An exact fulfillment of the Mermin-Wagner theorem is only expected when completely including $\Sigma_\infty\equiv\Sigma$ of Figs.~\ref{fig:mermin_wagner}(a) and (b).

\section{Definition of the step-width functions $s_x^y$ for $\Lambda$-integrations}
\label{app3}
The interval between consecutive RG steps $\Lambda_n$, $\Lambda_{n+1}$ in the numerical integration of the PFFRG equations is defined via
\begin{equation}
\Lambda_{n+1}=s(\Lambda_n)\Lambda_n\;.
\end{equation}
In Sec.~\ref{sec:results} we use different choices for the function $s(\Lambda_n)<1$ which are labeled by $s_x^y(\Lambda_n)$ and defined via
\begin{equation}
s_{0.995}^1(\Lambda_n)=\left\{\begin{array}{ll}0.95&\text{if }\Lambda_n>0.6\\0.98&\text{if } 0.6\geq\Lambda_n>0.4\\0.995&\text{if } 0.4\geq\Lambda_n\end{array}\right.\,,
\end{equation}
\begin{equation}
s_{0.995}^2(\Lambda_n)=\left\{\begin{array}{ll}0.95&\text{if }\Lambda_n>0.8\\0.98&\text{if } 0.8\geq\Lambda_n>0.6\\0.995&\text{if } 0.6\geq\Lambda_n\end{array}\right.\,,
\end{equation}
\begin{equation}
s_{0.995}^3(\Lambda_n)=\left\{\begin{array}{ll}0.95&\text{if }\Lambda_n>1.6\\0.98&\text{if } 1.6\geq\Lambda_n>1.2\\0.995&\text{if } 1.2\geq\Lambda_n\end{array}\right.\,.
\end{equation}
\vspace*{2cm}


%

\end{document}